\documentclass[aps,pra,twocolumn,superscriptaddress]{revtex4-2}
\usepackage{graphicx}
\usepackage{lipsum}
\usepackage{amsmath}
\usepackage{amssymb}
\usepackage{braket}
\usepackage{color}
\usepackage{dsfont}
\usepackage{gensymb}
\usepackage[hidelinks]{hyperref}
\usepackage{mathdots}
\usepackage{todonotes}
\renewcommand{\vec}[1]{\boldsymbol{#1}}
\begin{document}
\title{Beam focus and longitudinal polarization influence on spin dynamics in the Kapitza-Dirac effect}
\author{Sven Ahrens}
\email{ahrens@shnu.edu.cn}
\affiliation{Shanghai Normal University, Shanghai 200234, China}
\author{Ziling Guan}
\affiliation{Shanghai Normal University, Shanghai 200234, China}
\author{Baifei Shen}
\email{bfshen@shnu.edu.cn}
\affiliation{Shanghai Normal University, Shanghai 200234, China}
\date{\today}

\begin{abstract}
We theoretically investigate the influence of a longitudinal laser polarization component from beam focusing on spin dynamics in Kapitza-Dirac scattering by solving the relativistic Dirac equation with time-dependent perturbation theory. The transverse spacial dependence of the longitudinal beam polarization component is accounted for, by approximating a Gaussian beam with plane-wave components. We find that corrections from a longitudinal laser beam polarization component approximately scale with the second power of the diffraction angle $\epsilon$, from which we conclude that a related influence from beam focusing can be made negligibly small for sufficiently low beam foci.
\end{abstract}

\maketitle
\section{Introduction\label{sec:introduction}}
In recent years, theoretical investigations have suggested that the spin of an unbound electron in free space can be inferred by a standing wave of light
\cite{dellweg_mueller_2016_interferometric_spin-polarizer,dellweg_mueller_extended_KDE_calculations,ahrens_2017_spin_filter,ahrens_2020_two_photon_bragg_scattering}. The idea for the underlying electron diffraction effect in a standing light wave goes back to a proposal from Kapitza and Dirac in 1933 \cite{kapitza_dirac_1933_proposal} and with the discovery of the laser, first observation attempts were made in the 1960s \cite{schwarz_1965_KDE_dispute_1,pfeiffer_1968_KDE_dispute_2,takeda_1968_dispute_3}, which, however, were in dispute. Renewed attempts reported the observation of the
Kapitza-Dirac effect in 1980s for atoms in a strong interaction regime with many
diffraction orders \cite{gould_1986_atoms_diffraction_regime} and also in a weak interaction regime with isolated diffraction orders \cite{martin_1988_atoms_bragg_regime}. In the context of the Kapitza-Dirac effect, strong and weak interaction refer to a distinction between the diffraction regime (strong interaction), in which the energy-time uncertainty allows for multiple diffraction peaks, and the Bragg regime (weak interaction), where the duration of the interaction is typically sufficiently long, such that only one diffraction order is allowed \cite{batelaan_2000_KDE_first,batelaan_2007_RMP_KDE}. This one diffraction order in the Bragg regime only appears in a resonant configuration, where the diffracted particle and the absorbed and emitted laser photons need to fulfill the conservation of energy and momentum in the interpretation of a semiclassical interaction picture \cite{ahrens_bauke_2012_spin-kde,ahrens_bauke_2013_relativistic_KDE,ahrens_2012_phdthesis_KDE}. Subsequently, also Kapitza-Dirac scattering for electrons was observed in a high intensity interaction with many diffraction orders in 1988 \cite{Bucksbaum_1988_electron_diffraction_regime}. At the beginning of this century, in 2001, a rather precise setup for electrons with only a few diffraction orders has been carried out \cite{Freimund_Batelaan_2001_KDE_first}. This demonstration was followed by a refinement with only one diffraction order \cite{Freimund_Batelaan_2002_KDE_detection_PRL}, accordingly in the Bragg regime, which matches most the initial idea from Kapitza and Dirac.

With the experimental observation of the Kapitza-Dirac effect, the question arouse about whether Kapitza-Dirac scattering can also access the electron spin \cite{Batelaan_2003_MSGE}, where spin effects could not be reported for the considered scenario in reference \cite{Batelaan_2003_MSGE}, in an investigation based on classical particle trajectories. This motivated a quantum investigation on spin effects in the Kapitza-Dirac effect which was based on perturbative solutions of the Pauli equation in the diffraction regime, which also was not able to find pronounced spin effects \cite{rosenberg_2004_first_KDE_spin_calculation}. Ten years later, in 2012, a theoretical demonstration of significant spin effects in the Kapitza-Dirac effect was discussed within the context of a relativistic investigation of the Kapitza-Dirac effect \cite{ahrens_bauke_2012_spin-kde,ahrens_bauke_2013_relativistic_KDE}, where the change of the electron spin appears in resonant Rabi oscillations in the Bragg regime. The identification of resonant Rabi oscillations was inspired by similar resonances in the process electron positron pair-creation in counterpropagating laser beams \cite{ruf_2009_pair_creation}\footnote{We point out that even though the description of electron positron pair-creation demands for a many-particle context, the underlying formulation of pair-creation processes is related to solutions of the Dirac equation \cite{Fradkin_Gitman_Shvartsman_1991_Quantum_Electrodynamics_with_Unstable_Vacuum,woellert_2015_pair_creation_tunneling,woellert_2016_multi_pair_states,lv_bauke_2017_multi_pair_creation}. From this theoretical perspective, the difference of resonances in Bragg scattering in the Kapitza-Dirac effect and resonances in pair creation is only that pair creation is related to transitions from the negative to the positive energy continuum of the Dirac equation, whereas the electron resides at the positive energy-momentum dispersion relation for the case of the Kapitza-Dirac effect.}. Note, that one expects to approach the relativistic regime of the Kapitza-Dirac effect for electron momenta and also laser photon momenta larger than $1mc$, where for photon energies on the order of or larger than $1mc^2$ may also cause pair-creation processes. One may also expect relativistic effects for amplitudes of the vector potential $qA/(mc)>1$, as the electron may reach classical momenta larger than $1mc$. Nevertheless, one can show that spin dynamics are possible even in the non-relativistic regime, which can only be accounted for by relativistic corrections beyond the Pauli equation \cite{bauke_ahrens_2014_spin_precession_1,bauke_ahrens_2014_spin_precession_2}, ie. beyond the first order Foldy-Wouthuysen transformations \cite{foldy_wouthuysen_1950_non-relativistic_theory,greiner_2000_relativistic_quantum_mechanics}.

With indications for the possibility of spin interaction in the Kapitza-Dirac effect, further theoretical investigations in bichromatic standing light waves with frequency ratio 2:1 were carried out by using the Pauli equation \cite{McGregor_Batelaan_2015_two_color_spin,dellweg_awwad_mueller_2016_spin-dynamics_bichromatic_laser_fields,dellweg_mueller_2016_interferometric_spin-polarizer}. We mention that the authors in \cite{McGregor_Batelaan_2015_two_color_spin} also looked at classical electron trajectories based on the BMT equations, but found only vanishingly small spin-flip probabilities in the classical treatment. Also relativistic quantum calculations where made for bichromatic setups with the frequency ratio 2:1 \cite{dellweg_mueller_extended_KDE_calculations} and also for higher frequency ratios \cite{ebadati_2018_four_photon_KDE, ebadati_2019_n_photon_KDE}. The capability of spin-\emph{dependent} diffraction, in which the diffraction probability \emph{depends} on the initial electron spin state, appears as a novel property among most of the theoretical calculations of the bichromatic scenarios \cite{McGregor_Batelaan_2015_two_color_spin,dellweg_mueller_2016_interferometric_spin-polarizer,dellweg_mueller_extended_KDE_calculations,ebadati_2018_four_photon_KDE,ebadati_2019_n_photon_KDE}, where reference \cite{dellweg_mueller_2016_interferometric_spin-polarizer} demonstrates that this spin-dependent effect can also be achieved by using an interferometric setup.

Spin-dependent electron diffraction can also take place in monochromatic scenarios, in particular two-photon interactions for low electron momenta along the laser beam propagation direction \cite{ahrens_2017_spin_filter,ahrens_2020_two_photon_bragg_scattering}. While the spin-dependent effect in reference \cite{ahrens_2017_spin_filter} emerges only after the evolution of multiple Rabi cycles, reference \cite{ahrens_2020_two_photon_bragg_scattering} facilitates this effect already in the rise of the Bragg peak of the diffracted electron, which is beneficial for a possible experimental implementation with X-ray lasers. In the context of spin manipulations in laser-electron interactions, as discussed here, we also point out that the occurrence of electron spin polarization is discussed for ultra-relativistic laser-electron interactions \cite{PhysRevLett.123.174801,PhysRevLett.125.044802,PhysRevLett.122.154801,PhysRevLett.122.214801,PhysRevA.96.043407,PhysRevA.84.062116,article}.

The computation of the quantum dynamics in the Kapitza-Dirac effect is commonly carried out by assuming a plane wave laser field in most of the theoretical descriptions, where a final beam width and also a longitudinal polarization component from beam focusing of the laser are neglected. The question arises, whether the predicted spin effects are influenced by a beam with finite width or whether they are indeed negligible. We pick up this question in our article and compute the quantum dynamics of the Kapitza-Dirac effect with accounting for a small longitudinal polarization component from a Gaussian beam focus in a standing wave configuration. This longitudinal component would average to zero along the beam's transverse direction, such that we implement an additional transverse momentum degree of freedom in the electron wave function for the description of the diffraction process. A decomposition of the Gaussian laser field into an approximating superposition of plane waves allows us to still solve the problem analytically, within the framework of time-dependent perturbation theory.

Our article is organized as follows. In Sec. \ref{section II} we discuss the vector potential of the Gaussian beam and apply simplifying approximations to it for later calculations. After that, we introduce the Dirac equation in Sec. \ref{section III} and use it to establish a relativistic momentum space formulation of the quantum equations of motion, which are subsequently solved by time-dependent perturbation theory. The resulting propagation equation is then evaluated numerically in Sec. \ref{section IV}, from which we deduce a scaling behavior, which depends on the photon energy and the laser beam focusing angle. Finally, we discuss the influence of the longitudinal polarization component of the Gaussian beam on the electron spin dynamics in Sec. \ref{sec:discussion_and_conclusion} and list problems and potential future improvements of our description in the outlook in Sec. \ref{sec:outlook}.

\section{Setup and the vector potential of a Gaussian beam\label{section II}}

\subsection{Geometry of the investigated Kapitza-Dirac effect\label{sec:physical_setup}}

\begin{figure}%
    \includegraphics[width=0.5\textwidth]{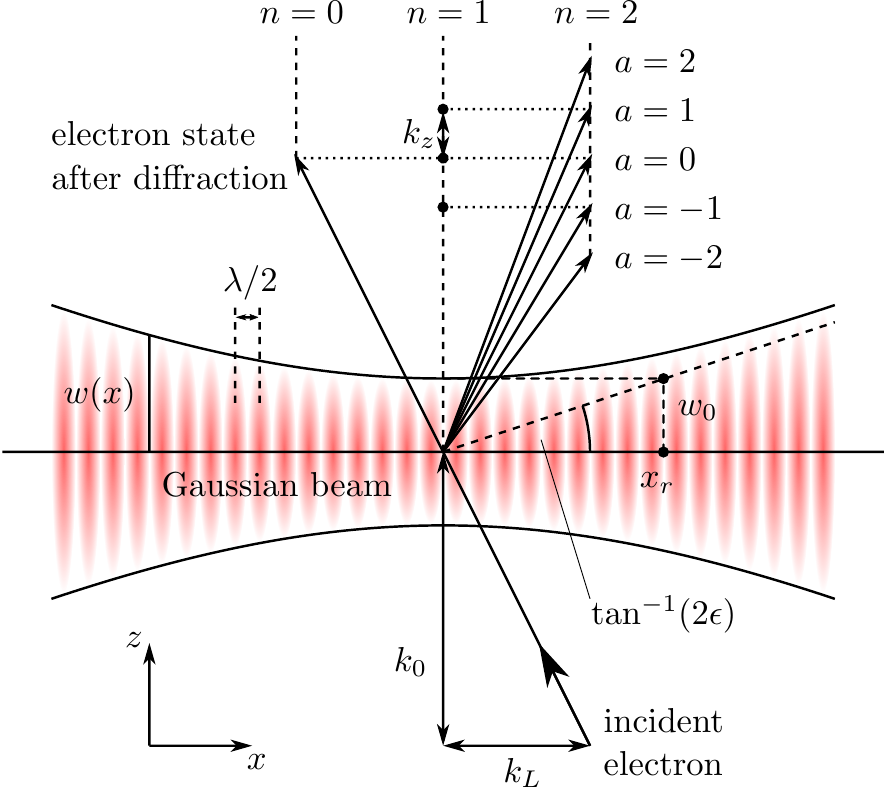}
    \caption{Geometric setup of the electron beam and the Gaussian standing wave laser beam. The Gaussian beam with wavelength $\lambda=2 \pi/k_L$, beam focus $w_0$, Rayleigh length $x_r$ and beam divergence $2 \epsilon$ is propagating along the $x$-axis. The electron beam is mainly propagating along the $z$-axis with $k_0=m\gg k_L$, where the electron momentum along the standing light wave is getting reversed on interaction with the laser in our investigated setup of the Kapitza-Dirac effect. To account also for the longitudinal beam polarization component, we consider a transverse momentum transfer in our approach with final diffraction orders $a \in \{-2, -1, 0, 1, 2\}$. The transverse momentum change in terms of multiples of momenta $k_z$ is smaller than the longitudinal momentum change, as implied by Eq. \eqref{eq:k_z_k_L_relation} and small beam divergences $2 \epsilon$. \label{fig:Gaussian_beam}}
\end{figure}%

The considered setup of our investigation is sketched in Fig. \ref{fig:Gaussian_beam}, in which the two counterpropagating laser beams of the standing light wave are propagating along the $x$-direction. Both beams are linearly polarized and the field of the vector potential is pointing in the $z$-direction. The laser beam has the wavelength $\lambda$ with corresponding wave number $k_L=2 \pi/\lambda$, beam waist $w_0$ at its focus, and the Rayleigh length $x_R=k_L w^2_{0}/2$. The quantity
\begin{equation}
 \epsilon=\frac{1}{k_L w_{0}}\,,\label{eq_epsilon_definition}
\end{equation}
as introduced in reference \cite{Quesnel_1998_gaussian_beam_coulomb_gauge} implies the ratio $w_{0}/x_R = 2 \epsilon$ and corresponds to the diffraction angle of the beam. For the momentum configuration of the electron, we follow previous investigations of such a laser setup \cite{ahrens_bauke_2013_relativistic_KDE,ahrens_2017_spin_non_conservation,ahrens_2020_two_photon_bragg_scattering}, in which spin effects occur for the transverse electron momentum $k_0=m$. Note, that we are using a Gaussian unit system with $\hbar=c=1$ in this article. Also, we use the words transverse ($z$-direction) and longitudinal ($x$-direction) with respect to the laser beam, if not stated differently. We also assume the system to be in the Bragg regime, which occurs for low
field amplitudes, and thus justifies the use of a perturbative technique for solving the quantum propagation of the electron. As mentioned in the introduction, the electron and the absorbed and emitted photons need to obey energy- and momentum conservation in the Bragg regime \cite{batelaan_2000_KDE_first,batelaan_2007_RMP_KDE}. From kinematic considerations \cite{ahrens_2012_phdthesis_KDE,ahrens_bauke_2012_spin-kde,ahrens_bauke_2013_relativistic_KDE} we know that this is only possible for initial and final electron momenta $\pm k_L \vec e_x$ along the $x$-axis, for the case of the monochromatic standing light wave which is considered here. In order to incorporate the longitudinal component of the Gaussian beam, it will also be necessary, to extend the plane wave expansion from a purely longitudinal degree of freedom for the electron momenta along the $x$-axis by adding a momentum degree of freedom along the transverse $z$-axis by multiples of momenta $k_z$. This becomes necessary for describing the non-negligible spacial $z$-dependence of the longitudinal potential \eqref{eq:Gaussain_beam_final_longitudinal_vector_potential} with the corresponding momentum space form \eqref{eq:S_piture_interaction_longitudinal_vector_potential}. In summary, the possible set of different electron momenta, which will appear in the extended plane wave ansatz \eqref{eq:the wave function of the Dirac equation}, are
\begin{equation}%
\vec k_{n,a}=(n-1)k_{L}\vec e_{x}+(ak_{z}+k_{0})\vec e_{z}\,. \label{eq:momentum_vector}
\end{equation}%
The Bragg condition, ie. the absorption and emission of one photon from each of the counterpropagating beams, implies that the electron is initially in a $n=0$ momentum state and finally in a $n=2$ momentum state. When transitioning from the $n=0$ to the $n=2$ state, a set of Kronecker deltas \eqref{D} will cause the initial transverse momentum state $a=0$ to be diffracted into a coherent superposition of momentum states $a\in\{-2,-1,0,1,2\}$ in our description. We have illustrated this form of superposition by five slightly diverging arrows, which are pointing from the origin towards the upper right in Fig. \ref{fig:Gaussian_beam}.

\subsection{Introduction of the vector potential of the Gaussian beam\label{sec:gaussian_beam_introduction}}

For the vector potential of the Gaussian beam we use a solution based on an angular spectrum representation of plane waves \cite{Quesnel_1998_gaussian_beam_coulomb_gauge}, which we write down in appendix \ref{sec:gaussian_vector_potential}, for completeness. After adjusting the solution to the desired geometry of our work, with a laser beam propagating along the $x$-axis, we obtain
\begin{subequations}%
\begin{align}%
A_{z,d}=&-A_{0}\frac{w_{0}}{w}\exp\left(-\frac{r^2}{w^2}\right)\sin\left(\phi_{G,d}\right)
\label{eq:transverse_vector_potential}
\end{align}%
for the transverse polarization component and
\begin{align}%
A_{x,d}=&-2dA_{0}\frac{w_{0}}{w}\epsilon\frac{z}{w}\exp\left(-\frac{r^2}{w^2}\right)\cos\left(\phi_{G,d}^{(1)}\right)
\label{eq:longitudinal_vector_potential}
\end{align}\label{eq:vector_field}%
\end{subequations}%
for the longitudinal polarization component of the vector potential of the Gaussian beam in Coulomb gauge. Eqs. \eqref{eq:vector_field} contain the two phases
\begin{subequations}%
\begin{align}%
\phi_{G,d}=&\omega t-dk_{L}x+\tan^{-1}\left(\frac{dx}{x_{R}}\right)-\frac{d x r^2}{x_{R}w^2}-\phi_{0,d}
\label{eq:add_equation1}\\
\phi_{G,d}^{(1)}=&\phi_{G,d}+\tan^{-1}\left(\frac{dx}{x_{R}}\right)\,.
\label{eq:add_equation2}
\end{align}\label{eq:Gaussian_beam_phase}%
\end{subequations}%
The symbol $A_0$ is the vector field amplitude and
\begin{equation}
 r = \sqrt{y^2 + z^2}
\end{equation}
“is the transverse distance from the beam propagation axis beam with $y=0$. We use the index $d$ to represent the direction of the beam, where $d \in\{-1,1\}$ corresponds to the left or right moving direction, respectively. The symbol $w$ is the $x$-dependent beam waist
\begin{equation}
w(x)=w_{0}\sqrt{1+\frac{x^2}{x_{R}^2}}\,,
\end{equation}
as illustrated in Fig. \ref{fig:Gaussian_beam}.  Note, that $A_x$  in Eq. \eqref{eq:transverse_vector_potential} is the additional longitudinal correction from beam focusing, which is of particular interest in this work. Since $A_x$ is proportional to $\epsilon$, it is getting vanishingly small for the case of arbitrary small beam foci.

\subsection{Application of approximations\label{sec:gaussian_beam_approximations}}

\begin{figure}%
    \includegraphics[width=0.5\textwidth]{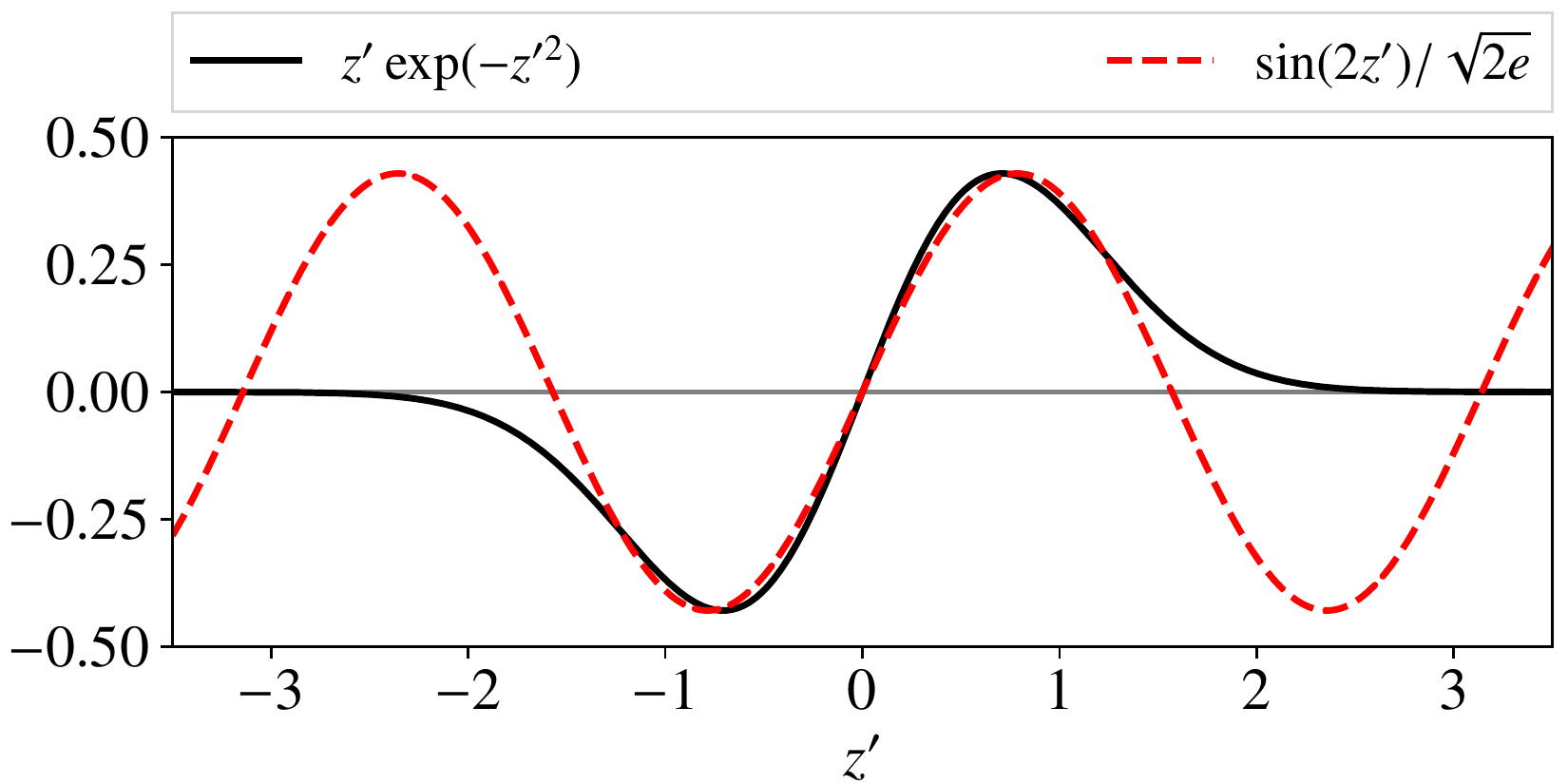}
    \caption{Illustration of the transverse beam dependence of the longitudinal polarization component \eqref{eq:longitudinal_z_dependence} (solid black line) and its approximating sine function (dashed red line). The sine function is inspired by two extrema of opposite sign at opposite locations around the origin in momentum space, as can be seen from the Fourier transform of Eq. \eqref{eq:longitudinal_z_dependence}, see main text. In position space, the sine function is chosen to match Eq. \eqref{eq:longitudinal_z_dependence} over the half period $-\pi/2 < z' < \pi/2$. The plot has been carried out over the reduced $z$-coordinate $z'=z/w$.\label{fig:gaussian_1_node}}
\end{figure}

In order to carry out the perturbative calculation in section \ref{section III}, it is necessary to simplify the potentials \eqref{eq:vector_field}, such that the expressions can be solved and written down. The longitudinal potential component \eqref{eq:longitudinal_vector_potential} would vanish, when simply averaged along the transverse direction. Therefore, the pure plane-wave ansatz as in previous calculations will not be capable of representing the influence of the longitudinal beam component. Instead we attempt the next possible increase of complexity of the description within a desired plane-wave like ansatz, which is capable of accounting for the longitudinal component. For the transverse component  \eqref{eq:transverse_vector_potential}, we desire the common plane wave approximation
\begin{equation}%
A_{z,d}=-A_{0}\sin\left(\phi_{G,d}\right)\,,\label{eq:simplified_transverse_plane_wave}
\end{equation}%
with the phase
\begin{equation}
 \phi_{G,d}=\omega t-dk_{L}x-\phi_{0,d}\,,\label{eq:plane_wave_phase}
\end{equation}
in place of Eq. \eqref{eq:add_equation1}. We desire a similarly simple form for the longitudinal component \eqref{eq:longitudinal_vector_potential}, where now we have to pay special attention to the odd (anti-symmetric) factor $z/w$, which causes the otherwise even (symmetric) function \eqref{eq:longitudinal_vector_potential} to vanish on average along the $z$-direction. On adopting the same phase in Eq. \eqref{eq:plane_wave_phase} also for $\phi_{G,d}^{(1)}$ in Eq. \eqref{eq:add_equation2}, we see that the only $z$-dependence in Eq. \eqref{eq:longitudinal_vector_potential} is given by
\begin{equation}
 \frac{z}{w}\exp\left(-\frac{z^2}{w^2}\right)\,.\label{eq:longitudinal_z_dependence}
\end{equation}
The Fourier transform and therewith functional form in momentum space of Eq. \eqref{eq:longitudinal_z_dependence} is $i p_z z w \exp[-(p_z w/2)^2]/\sqrt{8}$, with the conjugate $p_z$ of the $z$ variable. In the context of a simple approximation, the complex maximum at $0<p_z$ and complex minimum at $p_z<0$ can be represented by two spikes of delta functions with opposite signs, which constitute a sine function in position space. We display Eq. \eqref{eq:longitudinal_z_dependence} in Fig. \ref{fig:gaussian_1_node}, with the reduced $z$-coordinate $z'=z/w$.
The height of the extrema in position space is $1/\sqrt{2e}$ and with the argument $2 z'$, the approximating sine function matches Eq. \eqref{eq:longitudinal_z_dependence} over the range $-\pi/2 < z' < \pi/2$. Therefore, by imposing similar approximations as for the plane wave \eqref{eq:simplified_transverse_plane_wave} of the transverse polarization component also for the longitudinal polarization component \eqref{eq:longitudinal_vector_potential}, but also accounting for the odd $z$-dependence in Eq. \eqref{eq:longitudinal_z_dependence}, we simplify the longitudinal polarization component \eqref{eq:longitudinal_vector_potential} into
\begin{equation}
 A_{x,d}=-2dA_{0}\frac{\epsilon}{\sqrt{2 e}}\cos\left(\phi_{G,d}\right)\sin(z k_z)\,,\label{eq:simplified_longitudinal_plane_wave}
\end{equation}
where we introduce the transverse momentum displacement
\begin{equation}
k_{z}=\frac{2}{w_0}\,.\label{eq_k_z_definition}
\end{equation}
We mention that the definition for $\epsilon$ in Eq. \eqref{eq_epsilon_definition} and the specification for $k_z$ in Eq. \eqref{eq_k_z_definition} imply the relation
\begin{equation}
 k_z = 2 \epsilon k_L\,.\label{eq:k_z_k_L_relation}
\end{equation}
We also set $\phi_{0,d}=\pi$, for being consistent with the approach in reference \cite{ahrens_bauke_2013_relativistic_KDE}, resulting finally in
\begin{subequations}%
\begin{align}%
 A_{z,d}&=A_{0}\sin\left(\omega t-dk_{L}x\right)\label{eq:Gaussain_beam_transverse_vector_potential}\\
 A_{x,d}&=2dA_{0}\frac{\epsilon}{\sqrt{2 e}}\cos\left(\omega t-dk_{L}x\right)\sin(z k_z)\,.\label{eq:Gaussain_beam_longitudinal_vector_potential}
\end{align}\label{eq:final_plane_waves}%
\end{subequations}%

In this form, the approximated vector potential of the Gaussian beam is now suitable for conversion into a momentum space description with a manageable number of terms in Sec. \ref{sec:momentum_space_formulation} and carrying out the perturbative calculation in Sec. \ref{sec:time_dependent_perturbation_theory}. For ease of notion in subsequent calculations, we expand the trigonometric functions in Eq. \eqref{eq:final_plane_waves}: The sine part in Eq. \eqref{eq:Gaussain_beam_transverse_vector_potential} allows us to decompose the function $A_{z,d}$ into a sum of the exponential functions \\
\begin{align}
A_{z,d,o}=&-o\frac{i}{2}A_{0}e_{}^{oi(\omega t-dk_Lx)}
\label{eq:Gaussain_beam_final_transverse_vector_potential} 
\end{align}
where the index $o\in\{-1,1\}$ corresponds to either emission or absorption of a laser photon by the electron.\\ 

Correspondingly, the sine and cosine parts in Eq. \eqref{eq:Gaussain_beam_longitudinal_vector_potential} allow us to decompose the function $A_{x,d}$ into a sum of the exponential functions\\ 
\begin{align}
A_{x,d,o,f}=&-df\frac{i}{2}A_{0}\frac{\epsilon}{\sqrt[2]{2e}} e_{}^{oi(\omega t-dk_Lx)}e_{}^{fizk_{z}},
\label{eq:Gaussain_beam_final_longitudinal_vector_potential} 
\end{align}
with $o,f\in\{-1,1\}$, where $f$ corresponds to forward and backward motion of the electron along its propagation direction. We can therefore write Eqs. \eqref{eq:Gaussain_beam_transverse_vector_potential} and \eqref {eq:Gaussain_beam_longitudinal_vector_potential} as\\
\begin{subequations}%
\begin{align}%
A_{z,d}=\sum_{o}A_{z,d,o}\\
A_{x,d}=\sum_{o,f}A_{x,d,o,f}.
\end{align}\label{eq:potential_plane_wave_expansion}%
\end{subequations}%

\section{Theoretical description\label{section III}}

The approximated potentials \eqref{eq:Gaussain_beam_final_transverse_vector_potential} till \eqref{eq:potential_plane_wave_expansion} are consisting of plane waves, which will turn into Kronecker deltas when transforming them into the momentum space formulation \eqref{eq:S_piture_vector_potential}. This implies that only the subset of expansion coefficients $c_{n,a}^{\gamma,\sigma}(t)$ of the wave function's plane wave expansion \eqref{eq:the wave function of the Dirac equation} with the already introduced discrete momenta \eqref{eq:momentum_vector} are coupled to each other. Such a set of coefficients is suitable for applying the time-dependent perturbation theory calculation of section \ref{sec:time_dependent_perturbation_theory}, such that the result can be written down in a compact form. In order to see the emergence of the discrete Kronecker deltas in momentum space, one first needs to introduce the relativistic quantum description, on which the calculation is based. We therefore introduce the Dirac equation, which is done in the following section.

\subsection{The Dirac equation}

In quantum mechanics, the time-evolution of an electron  with mass $m$ and charge $q=-e$ is governed by\\
\begin{align}
i\frac{\partial \Psi(x)}{\partial t}=H \Psi(x),
\label{eq:derivation function}
\end{align}
where we aspire a relativistic quantum description with the Hamiltonian of the Dirac equation
\begin{align}
H=\alpha\left(\vec p-q \vec A\right)+qA_{}^{0}+\beta m\,.
\label{eq:the Dirac equation}
\end{align}
Here, we have introduced the $4\times 4$ Dirac matrices\\
\begin{align}%
\alpha_{i}=\begin{pmatrix}0&\sigma_{i}\\ \sigma_{i}&0\end{pmatrix},\beta=\begin{pmatrix}\mathds{1}&0\\0&-\mathds{1}\end{pmatrix}
\end{align}%
which contain the Pauli matrices\\
\begin{align}
\sigma_{x}=\begin{pmatrix}0&1\\1&0\end{pmatrix},\sigma_{y}=\begin{pmatrix}0&-i\\i&0\end{pmatrix},\sigma_{z}=\begin{pmatrix}1&0\\0&-1\end{pmatrix}
\label{eq:Pauli martrices}
\end{align}
\\
and the $2\times 2$ identity\\
\begin{align}
\mathds{1}=\begin{pmatrix}1&0\\0&1\end{pmatrix}.
\end{align}

\subsection{Momentum space formulation of the relativistic quantum theory\label{sec:momentum_space_formulation}}

The wave function $\Psi$ of the electron can be decomposed into a set of momentum and energy eigenfunctions
\begin{align}%
 \psi_{n,a}^{\gamma,\sigma}(\vec x)= &\sqrt{\frac{2\pi}{k_L}}\sqrt{\frac{2\pi}{k_z}}u_{\vec k_{n,a}}^{\gamma,\sigma}e_{}^{i\vec x \cdot \vec k_{n,a}},
\label{eq:the wave function}
\end{align}%
with the bi-spinors $u_{\vec k_{n,a}}^{\gamma,\sigma}$ defined as
\begin{subequations}%
\begin{align}%
u_{\vec k}^{+,\sigma}=\sqrt{\frac{E_{\vec k}+m}{2m}}
\begin{pmatrix}
\chi^\sigma\\ \frac{\vec \sigma \cdot \vec k}{E_{\vec k}+m}\chi^\sigma
\end{pmatrix}
\label{eq:the matrice 1}\\
u_{\vec k}^{-,\sigma}=\sqrt{\frac{E_{\vec k}+m}{2m}}
\begin{pmatrix}
- \frac{\vec \sigma \cdot \vec k}{E_{\vec k}+m}\chi^\sigma\\ \chi^\sigma
\end{pmatrix}.
\label{eq:the matrice 2}
\end{align}\label{eq:the matrice}%
\end{subequations}%
In Eqs. \eqref{eq:the wave function} and \eqref{eq:the matrice} the index $\gamma\in\{+,-\}$ denotes whether the electron is in a positive or negative energy eigenstate and the index $\sigma\in\{0,1\}$ denotes whether the electron is in a spin up $(0)$ or spin down $(1)$ state. The $n\in \mathbb{Z}$ index denotes the longitudinal momentum $(n-1)k_L$ of the electron beam in terms of laser photon momenta. The index $a$ denotes the transverse momentum $ak_z$ which is transferred from the transverse variation of the Gaussian beam's longitudinal component to the electron. 
Correspondingly, in Eq. \eqref{eq:the wave function} we are using the electron momentum \eqref{eq:momentum_vector}, resulting in the phase
\begin{equation}%
\vec x \cdot \vec k_{n,a}=(n-1)k_{L}x+(ak_{z}+k_{0})z \label{eq:the phase of the electron plane wave solution}
\end{equation}%
of the electron plane wave solution. The expression $E_{\vec{k}}$ is the relativistic energy-momentum relation
\begin{subequations}%
\begin{equation}%
E_{\vec{k}}=\sqrt{m_{}^{2}+k_{}^{2}}\,,
\end{equation}%
where we write%
\begin{equation}%
 E_{n,a} = \sqrt{m^2 + \vec k_{n,a}^2}
\end{equation}\label{eq:relativistic_energy_momentum_relation}%
\end{subequations}%
in place of $E_{\vec k}$ when using the discrete momenta $\vec k_{n,a}$. The variable $k_{0}$ parameterizes an initial transverse momentum of the electron along the $z$-axis.

With Eqs. \eqref{eq:the wave function} till \eqref{eq:relativistic_energy_momentum_relation}, we can write the wave function of the Dirac equation in momentum space as
\begin{align}%
\Psi(\vec x,t)=\sum_{\gamma,n,\sigma,a}c_{n,a}^{\gamma,\sigma}(t)\psi_{n,a}^{\gamma,\sigma}(\vec x).
\label{eq:the wave function of the Dirac equation}
\end{align}%
From this wave function expansion we denote the time-propagation of the initial expansion coefficients $c_{n,a}^{\gamma,\sigma}(t_0)$ into the the final expansion coefficients $c_{n',a'}^{\gamma',\sigma'}(t)$ for the plane-wave eigensolutions of the Dirac equation by\\
\begin{align}
c_{n',a'}^{\gamma',\sigma'}(t)=\sum_{\gamma,\sigma;n,a}U_{n',a';n,a}^{\gamma',\sigma';\gamma,\sigma}(t,t_0)c_{n,a}^{\gamma,\sigma}(t_0).
\label{eq:the possibility}
\end{align}
The approach in Eqs. \eqref{eq:the wave function} till \eqref{eq:the possibility} is extending similar formulations of the Dirac equation in momentum space \cite{ahrens_bauke_2012_spin-kde,ahrens_bauke_2013_relativistic_KDE,bauke_ahrens_2014_spin_precession_1,bauke_ahrens_2014_spin_precession_2,ahrens_2017_spin_filter,ahrens_2020_two_photon_bragg_scattering} by also introducing a transverse degree of freedom for the momentum of the electron wave function.

For the description of the quantum system by time-dependent perturbation theory we need a momentum space formulation of the interaction potentials. For this we denote the Dirac bra-ket notion
\begin{equation}
 \braket{\phi_a|Q|\phi_b} = \int d^3 x \, \phi_a^\dagger(\vec x) Q(\vec x) \phi_b(\vec x)\label{eq:dirac_bracket}
\end{equation}
of the matrix element $\braket{\phi_a|Q|\phi_b}$ for the operator $Q$. Based on this notion, we substitute the momentum eigenfunctions \eqref{eq:the wave function} into the two quantum states $\ket{\phi_a}$ and $\ket{\phi_b}$ and obtain the matrix elements
\begin{subequations}%
\begin{align}%
&V_{S;z,d,o,n',a';n,a\phantom{,f}}^{\gamma',\sigma' ;\gamma,\sigma}=\Braket{\psi_{n',a'}^{\gamma',\sigma'}|-qA_{z,d,o}\alpha_{3}|\psi_{n,a}^{\gamma,\sigma}}\nonumber\\
&=\frac{q}{2}oiA_{0}e_{}^{oi\omega t} L_{n',a';n,a,3}^{\gamma',\sigma' ;\gamma,\sigma}\delta_{a',a}\delta_{n',n-do} \label{eq:S_piture_interaction_transverse_vector_potential}\\
&V_{S;x,d,o,f,n',a';n,a}^{\gamma',\sigma' ;\gamma,\sigma}=\Braket{\psi_{n',a'}^{\gamma',\sigma'}|-qA_{x,d,o,f}\alpha_{1}|\psi_{n,a}^{\gamma,\sigma}}\nonumber\\
&=\frac{q}{2}dfiA_{0}e_{}^{oi\omega t}\frac{\epsilon}{\sqrt{2e}} L_{n',a';n,a,1}^{\gamma',\sigma' ;\gamma,\sigma}\delta_{a',a+f}\delta_{n',n-do}
\label{eq:S_piture_interaction_longitudinal_vector_potential}
\end{align}\label{eq:S_piture_vector_potential}%
\end{subequations}%
for the potentials $-qA_{z,d,o}\alpha_{3}$ and $-qA_{x,d,o,f}\alpha_{1}$, which include the expressions \eqref{eq:Gaussain_beam_final_transverse_vector_potential} and \eqref{eq:Gaussain_beam_final_longitudinal_vector_potential}. In Eqs. \eqref{eq:S_piture_vector_potential} we have introduced the abbreviation
\begin{align}\label{eq:S_piture_interaction_vector_potential}
L_{n',a';n,a;b}^{\gamma',\sigma' ;\gamma,\sigma}=\left(u_{\vec k_{n',a'}}^{\gamma',\sigma'}\right)_{}^{\dagger}\alpha_{b}\left(u_{\vec k_{n,a}}^{\gamma,\sigma}\right)\,.
\end{align}
The result in the second lines in Eqs. \eqref{eq:S_piture_vector_potential} is obtained by carrying out the space integration $\int d^3 x$ from the matrix element expression \eqref{eq:dirac_bracket}. For this integration, we denote all space dependent terms, which are the exponentials
\begin{subequations}%
\begin{equation}%
 \exp\left\{-i\left[\left(n'-n+do\right)k_L x + \left(a'-a\right)k_z z\right]\right\}\label{eq:transverse_phase}
\end{equation}%
for Eq. \eqref{eq:S_piture_interaction_transverse_vector_potential} and
\begin{equation}%
 \exp\left\{-i\left[\left(n'-n+do\right)k_L x + \left(a'-a-f\right)k_z z\right]\right\}\label{eq:longitudinal_phase}
\end{equation}\label{eq:potential_phases}%
\end{subequations}%
for Eq. \eqref{eq:S_piture_interaction_longitudinal_vector_potential}. When carrying out the three dimensional integration, the phases collapse into the Kronecker deltas $\delta_{a',a}\delta_{n',n-do}$ (in Eq. \eqref{eq:S_piture_interaction_transverse_vector_potential}, originating from Eq. \eqref{eq:transverse_phase}) and $\delta_{a',a+f}\delta_{n',n-do}$ (in Eq. \eqref{eq:S_piture_interaction_longitudinal_vector_potential}, originating from Eq. \eqref{eq:longitudinal_phase}).

One can similarly obtain the momentum space formulation of the Dirac equation
\begin{align}
i\hbar\frac{\partial}{\partial t}c=Ec+\sum V_{s,z}c +\sum V_{s,x}c
\label{eq:the sum possibility}
\end{align}
by projecting the adjoint plane wave solutions \eqref{eq:the wave function} from the left on the time-evolution equation \eqref{eq:derivation function}, as done already in references \cite{ahrens_bauke_2012_spin-kde,ahrens_bauke_2013_relativistic_KDE,ahrens_2017_spin_filter,ahrens_2020_two_photon_bragg_scattering}. Note, that in Eq. \eqref{eq:the sum possibility}, we have omitted the indices and time-dependence for the expansion coefficients $c^{\gamma,\sigma}_{n,a}(t)$ and the potentials \eqref{eq:S_piture_vector_potential} in favor of a compact notion. Still, the sums in Eq. \eqref{eq:the sum possibility} run over the unprimed indices, as they appear in Eqs. \eqref{eq:S_piture_vector_potential}. The expansion coefficients on the left-hand side and the first term on the right-hand side of Eq. \eqref{eq:the sum possibility} have primed indices, ie.  $c^{\gamma',\sigma'}_{n',a'}(t)$. The symbol $E$ denotes the relativistic energy-momentum relation \eqref{eq:relativistic_energy_momentum_relation}, which can be positive and negative, corresponding to the expansion coefficients $c^{\gamma',\sigma'}_{n',a'}$ of the positive and negative energy eigensolutions. We will make use of the shortened notion in Eq. \eqref{eq:the sum possibility} with omitted indices and omitted time-dependence also in subsequent expressions of similar complexity in the remaining text of this article.

\subsection{Time-dependent perturbation theory\label{sec:time_dependent_perturbation_theory}}

In order to calculate the time-evolution of the quantum state, we are making use of second order time-dependent perturbation theory \cite{sakurai2014modern}\\
\begin{align}
U(t,t_{0})=(-i)^2\int_{t_{0}}^{t}dt_{2}\int_{t_{0}}^{t_{2}}dt_{1}V_{I}(t_{2})V_{I}(t_{1})\,,
\label{eq:time pertubation theory of V}
\end{align}
where we follow the convention to carry out our calculation in the interaction picture, with operators related by\\
\begin{align}
V_{I}=e_{}^{iH_{0}t}V_{S}e_{}^{-iH_{0}t}\,.
\label{picture change_a}
\end{align}
Here, $V_S$ and $V_I$ are the operators in the Schr\"odinger and interaction picture, respectively. With the matrix elements $\gamma E_{\vec k_{n,a}}$ of the free Hamiltonian $H_0$ in momentum space, relation \eqref{picture change_a} becomes
\begin{align}
V_{I;n',a';n,a}^{\gamma',\sigma';\gamma,\sigma}=V_{S;n',a';n,a}^{\gamma',\sigma' ;\gamma,\sigma}e_{}^{i(\gamma' E_{n',a'}-\gamma E_{n,a})t}
\label{picture change_b}
\end{align}
in explicit index notation. By inserting the potentials \eqref{eq:S_piture_vector_potential} of the interaction picture \eqref{picture change_b} into the perturbation expression \eqref{eq:time pertubation theory of V}, we obtain
\begin{align}
U(t,t_{0})=-\sum\Gamma\frac{q^2A_{0}^2}{4}D\Lambda\int_{t_{0}}^{t}dt_{2}\int_{t_{0}}^{t_{2}}dt_{1}\Upsilon
\label{eq:time perturbation theory}\,.
\end{align}
In Eq. \eqref{eq:time perturbation theory} the expression $D$ is a collection of Kronecker deltas
\begin{subequations}%
\begin{align}%
D_{z,z}=&\delta_{n'',n'-d'o'}\delta_{n',n-do}\delta_{a'',a'}\delta_{a',a}\\
D_{x,z}=&\delta_{n'',n'-d'o'}\delta_{n',n-do}\delta_{a'',a'+f'}\delta_{a',a}\\
D_{z,x}=&\delta_{n'',n'-d'o'}\delta_{n',n-do}\delta_{a'',a'}\delta_{a',a+f}\\
D_{x,x}=&\delta_{n'',n'-d'o'}\delta_{n',n-do}\delta_{a'',a'+f'}\delta_{a',a+f}\,,
\end{align}\label{D}%
\end{subequations}%
which originates from the potentials \eqref{eq:S_piture_vector_potential}. The expression $\Lambda$ is the corresponding collection of the spin-dependent terms
\begin{align}
\Lambda_{n'',a'';n',a';n,a;r,t}^{\gamma'',\sigma'';\gamma',\sigma';\gamma,\sigma}=L_{n'',a'';n',a';r}^{\gamma'',\sigma'';\gamma',\sigma'}L_{n',a';n,a;t}^{\gamma',\sigma';\gamma,\sigma}\,.
\label{the collection of spin-dependent terms}
\end{align}
All time-dependent expressions have been absorbed in the time-dependent phase
\begin{multline}%
\Upsilon_{n'',a'';n',a',n,a}^{\gamma'',\gamma',\gamma;o',o}(t_{2},t_{1})= \\ e_{}^{i(\gamma ''E_{n'',a''}-\gamma'E_{n',a'}+o'\omega)t_{2}} \\ \times e_{}^{i(\gamma 'E_{n',a'}-\gamma E_{n,a}+o\omega)t_{1}}\label{the time-dependent phase from the interaction}%
\end{multline}%
behind the final, double-time integral $\int_{t_{0}}^{t}dt_{2}\int_{t_{0}}^{t_{2}}dt_{1}$. All other prefactors, which cannot be summarized in a simple way, are combined in the prefactor $\Gamma$ as
\begin{subequations}%
\begin{align}%
\Gamma_{z,z}&=oo'\\
\Gamma_{x,z}&=d'f'o\mathcal{E}\\
\Gamma_{z,x}&=dfo'\mathcal{E}\\
\Gamma_{x,x}&=d'df'f\mathcal{E}^2\,,
\end{align}\label{prefactor}%
\end{subequations}%
where the calligraphically written $\mathcal{E}$ is an abbreviation for the scaled diffraction angle $\epsilon$
\begin{align}
\mathcal{E}=\frac{\epsilon}{\sqrt{2e}}\,.\label{eq:capital_epsilon}
\end{align}
The index pairs $\{(z,z);(x,z);(z,x);(x,x)\}$, which we have attached to $D$ and $\Gamma$ are accounting on whether $V_z$ (Eq. \eqref{eq:S_piture_interaction_transverse_vector_potential}) or $V_x$ (Eq. \eqref{eq:S_piture_interaction_longitudinal_vector_potential}) have been used for the potential $V_I(t_2)$ in Eq. \eqref{eq:time pertubation theory of V} (first index $t_2$) and on whether $V_z$ or $V_x$ have been used for $V_I(t_1)$ (second index $t_1$).\\
 Note, that in Eq. \eqref{eq:time perturbation theory}, we have omitted the indices for the expansion coefficients $U_{n'',a'';n,a}^{\gamma'',\sigma'';\gamma,\sigma}$ in a similar way as we have done it for Eq. \eqref{eq:the sum possibility}. Correspondingly, the sum in Eq. \eqref{eq:time perturbation theory} runs over the indices $\gamma'$, $\sigma'$ , $n'$ and $a'$ as part of the matrix product between the potentials \eqref{eq:S_piture_vector_potential}. Additionally, the sum also runs over the possible configurations $o$, $o'$, $d$ and $d'$.
 
\subsection{The resonance condition in the Bragg regime of the Kapitza-Dirac effect}
We proceed the computation of the perturbative expression \eqref{eq:time perturbation theory} by solving the double time integral $\int_{t_{0}}^{t}dt_{2}\int_{t_{0}}^{t_{2}}dt_{1}\Upsilon$. The integral $\int_{t_{0}}^{t_{2}}dt_{1}$ over the $t_1$ dependent exponential in \eqref{the time-dependent phase from the interaction} results in
\begin{multline}
 \int_{t_{0}}^{t_{2}}dt_{1} e_{}^{i(\gamma' E_{n',a'}-\gamma E_{n,a}+o\omega)t_{1}}\\
 = i F \left. e_{}^{i(\gamma' E_{n',a'}-\gamma E_{n,a}+o\omega)t_{1}} \right|_{t_0}^{t_2}\,,\label{eq:t_1_integration}
\end{multline}
where we have introduced the abbreviation
\begin{align}
F=(\gamma E_{n,a}-\gamma' E_{n',a'}-o\omega)_{}^{-1}\,.
\label{F}
\end{align}
For the upper integration limit $t_2$ in Eq. \eqref{eq:t_1_integration} we obtain
\begin{equation}
 i F \int_{t_{0}}^{t}dt_{2} e_{}^{i(\gamma ''E_{n'',a''}-\gamma E_{n,a}+o'\omega+o\omega)t_{2}}\label{eq:t2_integral}
\end{equation}
in the double integral $\int_{t_{0}}^{t}dt_{2}\int_{t_{0}}^{t_{2}}dt_{1}\Upsilon$. The argument in the exponent, which we abbreviate by
\begin{equation}
 \Delta E = \gamma ''E_{n'',a''}-\gamma E_{n,a}+o'\omega+o\omega\,,\label{eq:delta_E}
\end{equation}
corresponds to the net energy transfer of the two interacting laser photons with the electron. With Eq. \eqref{eq:delta_E} the solution of \eqref{eq:t2_integral} can be written as
\begin{subequations}%
\begin{align}%
 &i F \int_{t_{0}}^{t}dt_{2} e_{}^{i \Delta E t_{2}} = \frac{F}{\Delta E} \left( e^{i \Delta E t} - e^{i \Delta E t_0 } \right)\label{eq:expanded_phase_integral_oscillating}\\
 &\qquad = i F \sum_{g=0}^\infty \frac{(i \Delta E)^g}{(g+1)!} \left(t^{g+1} - t_0^{g+1}\right)\\
 &\qquad = i F \left[ t-t_0 + \frac{i \Delta E}{2} \left(t^2 - t_0^2\right) + \dots \right]\,.\label{eq:expanded_phase_integral_explicitly}
\end{align}\label{eq:expanded_phase_integral}%
\end{subequations}%
As explained in the introduction, Kapitza-Dirac scattering takes place for 
\begin{equation}
 n=0\,, \qquad n=2 \,,\label{eq:x_momentum_constraint}
\end{equation}
for the positive particle solutions
\begin{equation}
 \gamma=\gamma^{\prime\prime}=+1\,,
\end{equation}
with one absorbed and one emitted photon, corresponding to
\begin{equation}
 o=-o'\in \{-1,1\}\,,\label{eq:photon_absorption_emission}
\end{equation}
see references \cite{ahrens_2012_phdthesis_KDE,ahrens_bauke_2012_spin-kde,ahrens_bauke_2013_relativistic_KDE} for details. In the case of no transverse momentum transfer, ie. $a=a^{\prime\prime}=0$, one sees that $\Delta E$ in Eq. \eqref{eq:delta_E} vanishes, such that the solution \eqref{eq:expanded_phase_integral_explicitly} of the upper limit of the integral over $t_1$ in the double time integral of Eq. \eqref{eq:time perturbation theory} simplifies to
\begin{align}
\int_{t_{0}}^{t}dt_{2}\int^{t_{2}}dt_{1}\Upsilon(t_2,t_1)= i F(t-t_0)\,.
\label{double-time integral}
\end{align}
In this resonant situation, in which the phase of the incoming and outgoing mode of the electron wave function are in resonance with the phase oscillations of the interacting photons, the amplitude of the diffracted mode can grow unlimited in time. In the case of a perfect resonant situation with $\Delta E = 0$, this growth would be unbound and only constrained by the unitary property of the Dirac equation, where this unitary property in turn would manifest itself only in higher order perturbative contributions (as an example see the calculation of the Kapitza-Dirac effect based on the Schr\"odinger equation in reference \cite{gush_gush_1971_higher_order_kapitza_dirac_scattering}). One can see in Eq. \eqref{eq:expanded_phase_integral} that the resonant mode, which grows linear with the interaction time $\Delta t = t - t_0$ can outgrow the oscillating solution \eqref{eq:expanded_phase_integral_oscillating}, when the product $\Delta E \Delta t$ is approximately smaller than one. This recovers the energy-time uncertainty condition, which is used to distinguish between diffraction regime ($\Delta E \Delta t$ larger one) and Bragg regime ($\Delta E \Delta t$ smaller one), according to Batelaan \cite{batelaan_2000_KDE_first,batelaan_2007_RMP_KDE} (see also the discussion in the introductory section \ref{sec:introduction}). Corresponding resonance peaks of the diffraction amplitude which illustrate this energy-time uncertainty are shown, for example, in references \cite{ahrens_2012_phdthesis_KDE,ahrens_bauke_2013_relativistic_KDE,dellweg_awwad_mueller_2016_spin-dynamics_bichromatic_laser_fields}.

For our investigation we will assume the dynamics to be on resonance, with $\Delta E=0$ in \eqref{eq:delta_E}, as the integration result \eqref{double-time integral} will outgrow all other oscillatory contributions in the integral $\int_{t_{0}}^{t}dt_{2}\int_{t_{0}}^{t_{2}}dt_{1}\Upsilon$. Also, since according to Eq. \eqref{eq:k_z_k_L_relation} the transverse momentum transfer $k_z$ is smaller than the longitudinal momentum transfer $k_L$ by the factor $2 \epsilon$, where $\epsilon$ is usually much smaller than one, we also assume that diffraction into the final states with $a^{\prime\prime} \in \{-2, -1, 0, 1, 2\}$ are also on resonance, ie. $E_{n^{\prime\prime},a^{\prime\prime}}$ being very close to $E_{n,a}$. We thus assume to obtain the result \eqref{double-time integral}, independently from the value of $a^{\prime\prime}$. 

We also point out that the absolute value of the momentum transfer $k_L \vec e_x \pm k_z \vec e_z$ which is implied by the longitudinal potential \eqref{eq:S_piture_interaction_longitudinal_vector_potential} is larger than the absolute value of the corresponding momentum transfer $k_L \vec e_x$ of the transverse potential \eqref{eq:S_piture_interaction_transverse_vector_potential}, as a result of the approximation in section \ref{sec:gaussian_beam_approximations}. We are therefore using the vacuum dispersion relation of light
\begin{subequations}%
\begin{equation}%
 \omega_{n,a;n',a'} = \left| \vec k_{n,a} - \vec k_{n',a'} \right|\label{eq:photon_vacuum_dispersion}
\end{equation}%
for the prefactor in Eq. \eqref{F}, which we explicitly write as
\begin{align}%
F=(\gamma E_{n,a}-\gamma' E_{n',a'} - o \, \omega_{n,a;n',a'})^{-1}\,.\label{eq:F_improved}
\end{align}\label{eq:F_with_vacuum_dispersion}%
\end{subequations}%
Using a only a constant dispersion $\omega = k_L$ would result in situations in which the bracket on the right-hand side of \eqref{eq:F_improved} goes to zero and causes the diffraction amplitude to diverge. Such a divergence only takes place for unphysically large the beam divergence angles, at which $k_z \gtrapprox k_L$. For this reason we consider the implementation of the vacuum dispersion \eqref{eq:F_with_vacuum_dispersion} as appropriate.

We finally can write the expression for the perturbative calculation as
\begin{align}
U(t,t_{0})=- i\sum\Gamma F\frac{q^2A_{0}^2}{4}\Lambda(t-t_{0})\,,
\label{eq:solution_perturbation_theory}
\end{align}
where we have substituted the solution of the double time integral \eqref{double-time integral} with the prefactor \eqref{eq:F_with_vacuum_dispersion} from the integration into the intermediate perturbative expression \eqref{eq:time perturbation theory}.

\subsection{Momentum conservation}

In order to complete the perturbative calculation, the electron momenta in Eq. \eqref{eq:solution_perturbation_theory} need to be specified, still. For that, we make use of the momentum conservation, which is implied by the Kronecker deltas in Eq. \eqref{D}. The $x$-dependent Kronecker deltas with dependence of $n$, $n'$ and $n''$ imply the conditions
\begin{subequations}%
\begin{align}%
 n' &=n - do\\
 n''&=n' - d'o'\,.
\end{align}\label{eq:x_momentum_conservation_separate}%
\end{subequations}%
To resolve this, we refer back to the initial and final $x$-momentum constraints \eqref{eq:x_momentum_constraint} and the photon absorption and emission condition \eqref{eq:photon_absorption_emission}. We first note that reaching from $n=0$ to $n''=2$ is only possible for $n'=1$. Secondly, combining the two conditions in \eqref{eq:x_momentum_conservation_separate} results in
\begin{equation}
 n'' = n - d' o' - d o\,.\label{eq:x_momentum_conservation}
\end{equation}
For the defined range of the parameters $o, o', d, d' \in \{-1,1\}$, the conditions \eqref{eq:x_momentum_constraint}, \eqref{eq:photon_absorption_emission} and \eqref{eq:x_momentum_conservation} impose the condition
\begin{equation}
 d = - o = o' = -d'\,. \label{od_index_fix}
\end{equation}
For the transverse direction ($z$-direction) we require the electron to move with momentum $k_0$, corresponding to $a=0$, as implied by the approach for the electron momentum in Eq. \eqref{eq:momentum_vector}. The $z$-dependent Kronecker deltas in Eq. \eqref{D} with dependence of $a$, $a'$ and $a''$ imply the conditions
\begin{subequations}%
\begin{align}%
 a' &=a  + f\\
 a''&=a' + f'\,,
\end{align}\label{eq:transverse_momentum_conservation}%
\end{subequations}%
where occurrences of $\delta_{a',a}$ and $\delta_{a'',a'}$ can be accounted for, in the form \eqref{eq:transverse_momentum_conservation}, by setting $f=0$ and $f'=0$, respectively. With Eqs. \eqref{eq:transverse_momentum_conservation} we can determine the $\Gamma$ factors in Eq. \eqref{prefactor} for different values of $a'$ and $a''$, if we additionally make use of the implications $oo'=dd'=-1$ and $od'=do'=1$ from Eq. \eqref{od_index_fix}. All possible combinations for $\Gamma$ are listed in table \ref{tabel I}.

\begin{table}
\caption{
Specific values of the polarization dependent $\Gamma$ prefactor \eqref{prefactor}, as it appears in the perturbative expression \eqref{eq:solution_perturbation_theory}. The electron quantum state propagation happens on different quantum trajectories in momentum space, where the electron can be subject to different polarization components when interacting with the laser (see main text below Eq. \eqref{eq:capital_epsilon}). The index pair $(z,z)$ scales with zero power in $\mathcal{E}$, the index pairs $(x,z)$ and $(z,x)$ scale with one power in $\mathcal{E}$ and $(z,z)$ scales with two powers in $\mathcal{E}$. We have separated the different powers of $\mathcal{E}$ with double lines. We also have separated different diffraction orders $a''$ of the electron's final momentum $a'' k_z + k_0$ along the transverse laser direction by additional, single lines.\label{tabel I}}

\begin{tabular}{ r  r  l }			
  $a''$ & $a'$ & $\Gamma$\\
  \hline \hline
  0 & 0 & $\Gamma_{z,z}=-1$ \\
  \hline \hline
  1 & 0 & $\Gamma_{x,z}=+\mathcal{E}$\\
  1 & 1 & $\Gamma_{z,x}=+\mathcal{E}$\\
  \hline  
  -1 & 0 & $\Gamma_{x,z}=-\mathcal{E}$\\
  -1 & -1 & $\Gamma_{z,x}=-\mathcal{E}$ \\
  \hline \hline
   2 & 1 & $\Gamma_{x,x}=-\mathcal{E}^2$\\
  \hline
   0 & 1 & $\Gamma_{x,x}=+\mathcal{E}^2$\\
   0 & -1 & $\Gamma_{x,x}=+\mathcal{E}^2$\\
   \hline
  -2 & -1 & $\Gamma_{x,x}=-\mathcal{E}^2$\\
\end{tabular}
\end{table}

The role of all indices in the final perturbative expression \eqref{eq:solution_perturbation_theory} are determined with the above considerations and they can be classified into the four different categories:\\
Indices with fixed values:\\
{\color{white}.}\hspace{0.5 cm}$n=0$, $n'=1$, $n''=2$, $a=0$, $\gamma=1$, $\gamma''=1$\\
Indices which still appear in the sum of Eq. \eqref{eq:solution_perturbation_theory}:\\
{\color{white}.}\hspace{0.5 cm}$a'$, $\sigma'$, $\gamma'$, $o$\\
Indices which are implied by Eqs. \eqref{od_index_fix} and \eqref{eq:transverse_momentum_conservation}:\\
{\color{white}.}\hspace{0.5 cm}$o'$, $d$, $d'$, $f$, $f'$\\
Indices, which are not determined yet:\\
{\color{white}.}\hspace{0.5 cm}$a''\in\{-2,-1,0,1,2\}$, $\sigma$, $\sigma''$

\section{Resulting modification of spin-preserving and spin-changing terms\label{section IV}}

\subsection{Investigation procedure}

In this section we want to quantify the influence of the longitudinal polarization component from the Gaussian beam on the spin dynamics in the Kapitza-Dirac effect.  Since the perturbative expression \eqref{eq:solution_perturbation_theory} in section \ref{section III} has a complicated structure, we want to investigate its dependence on the photon energy $k_L$ and the transverse momentum transfer $k_z$ numerically. To do that, we start in section \ref{sec:spin_propagation} with first denoting a formalism for decomposing the quantum state propagation into spin-preserving and spin-changing components, which can be seen in Eqs. \eqref{eq:spin-conserved} and \eqref{eq:spin-changing}. This makes it more easy to identify the influence of the longitudinal beam component on the spin dynamics. The formalism also has the advantage that it is independent of the initial and final electron spin configuration. We then want to numerically extract simple power law scaling relations for \eqref{eq:propagator_projections}, in the form of Eq. \eqref{eq:power_law_function}. We do that, by first plotting the functional dependence of \eqref{eq:propagator_projections} as a function of $k_L$ and/or $k_z$ in Figs. \ref{Fig.1} till \ref{Fig.4} in section \ref{sec:numeric_evaluation}. The figures are carried out as double logarithmic plots, such that power law scalings appear as a straight lines which can be fitted to linear functions, to obtain the coefficients of the power law functions \eqref{eq:power_law_function}. This is done in section \ref{sec:analysis_of_results} and results in the scaling functions \eqref{eq:scaling_longitudinal_1} till \eqref{eq:scaling_longitudinal_2}. Finally, the obtained scaling relations are then compared with the corresponding expression in which no longitudinal polarization component has been used, to obtain relation \eqref{eq:beam_focus_relevance}, from which one can see for what parameters $k_L$ and $\epsilon$ the longitudinal polarization component from beam focusing becomes relevant.

\subsection{Spin propagation\label{sec:spin_propagation}}

The initial electron spin configuration $c_{0,0}^{1,\sigma}(t_0)$ is diffracted by the laser interaction into the final electron spin configuration $c_{2,a''}^{1,\sigma''}(t)$ by
\begin{equation}
 c_{2,a''}^{1,\sigma''}(t)=\sum_{\sigma} U_{2,a'';0,0}^{1,\sigma'';1,\sigma}(t,t_0)c_{0,0}^{1,\sigma}(t_0)\label{eq:spin_propagation}
\end{equation}
in terms of the general quantum state propagation equation \eqref{eq:the possibility}. The entries of
\begin{equation}
U_{2,a'';0,0}^{1,\sigma'';1,\sigma}(t,t_0) =
\begin{pmatrix}
 u_{00} & u_{01} \\
 u_{10} & u_{11}
\end{pmatrix}
\end{equation}
for a specific value of $a''$ are then the entries of a complex $2 \times 2$ matrix with column index $\sigma''$ and row index $\sigma$. In the abstract 4-component vector space $(u_{00}, u_{01}, u_{10}, u_{11})^T \in \mathbb{C}^4$ of complex $2 \times 2$ matrices, one can define the scaled inner product
\begin{align}
\Braket{M|U}=\frac{1}{\eta} \left(m_{00}^{*}u_{00}+m_{01}^{*}u_{01}+m_{10}^{*}u_{10}+m_{11}^{*}u_{11}\right)\,,
\label{eq:40}
\end{align}
with $u, m \in \mathbb{C}^{2\times 2}$ being the matrix entries of the corresponding $2 \times 2$ matrices $U$ and $M$, respectively. A scale parameter $\eta$ appears in \eqref{eq:40} which will be specified soon in Eq. \eqref{eq:eta_definition}. The space of complex $2\times 2$ matrices can be spanned by the $2\times 2$ identity matrix $\mathds{1}$ and the three Pauli matrices $\sigma_x$, $\sigma_y$, $\sigma_z$. Projecting the numerically evaluated expressions of $U_{2,a'';0,0}^{1,\sigma'';1,\sigma}(t,t_0)$ in the form of Eq. \eqref{eq:solution_perturbation_theory} on these four orthogonal components yields only non-vanishing contributions for $\Braket{\mathds{1}|U}$ and $\Braket{\sigma_{y}|U}$ for the field configuration of the combined fields \eqref{eq:Gaussain_beam_transverse_vector_potential} and \eqref{eq:Gaussain_beam_longitudinal_vector_potential}. In terms of the inner product notion \eqref{eq:40}, we therefore set
\begin{subequations}%
\begin{align}%
\Braket{\mathds{1}|U}_{a''}&=\frac{1}{\eta}\left[U_{2,a'';0,0}^{1,0;1,0}(t,t_{0})+U_{2,a'';0,0}^{1,1;1,1}(t,t_{0})\right]\label{eq:spin-conserved}
\\
\Braket{\sigma_{y}|U}_{a''}&=\frac{i}{\eta}\left[U_{2,a'';0,0}^{1,0;1,1}(t,t_{0})-U_{2,a'';0,0}^{1,1;1,0}(t,t_{0})\right]\,,\label{eq:spin-changing}
\end{align}\label{eq:propagator_projections}%
\end{subequations}%
where we factor out the value
\begin{equation}
 \eta=2\left[-i \,\Gamma \frac{q^2A_{0}^2}{4}(t-t_0)\right]\,.\label{eq:eta_definition}
\end{equation}
In this way, $U$ appears in the form
\begin{equation}
\begin{pmatrix}
U_{2,a'';0,0}^{1,0;1,0} & U_{2,a'';0,0}^{1,0;1,1}\\
U_{2,a'';0,0}^{1,1;1,0} & U_{2,a'';0,0}^{1,1;1,1}
\end{pmatrix}
= \frac{\eta}{2}
\begin{pmatrix}
  \Braket{\mathds{1}|U}_{a''} & -i \Braket{\sigma_{y}|U}_{a''}\\
i \Braket{\sigma_{y}|U}_{a''} &    \Braket{\mathds{1}|U}_{a''}
\end{pmatrix}\,.\label{eq:spin_propagation_matrix}
\end{equation}
We also find numerically that $\Braket{\mathds{1}|U}$ is purely real and $\Braket{\sigma_{y}|U}$ is purely imaginary, ie.
\begin{subequations}%
\begin{align}%
 \textrm{Im}(\Braket{\mathds{1}|U})&=0\\
 \textrm{Re}(\Braket{\sigma_{y}|U})&=0\,,
\end{align}\label{eq:spin_decomposition_real_and_imaginary_value}%
\end{subequations}%
for each index $a''$. In the following, we want to give a more intuitive picture of the spin decomposition in this subsection, regarding the physical point of view of our description.

Based on the property \eqref{eq:spin_decomposition_real_and_imaginary_value}, one can further substitute
\begin{subequations}%
\begin{align}%
 \Braket{\mathds{1}|U} &= \xi \cos \frac{\phi}{2} \\
 \Braket{\mathds{\sigma}_y|U} &= -i \xi \sin \frac{\phi}{2}\,,
\end{align}%
\end{subequations}%
with an amplitude $\xi$ and an angle $\phi$, for each index $a''$. In this form Eq. \eqref{eq:spin_propagation_matrix} turns into a $\mathcal{SU}(2)$ matrix times an amplitude $\eta \xi/2$ and acts at the quantum state as a spin rotation, combined with a diffraction probability. This can be seen by assuming the initial electron quantum state $c_{0,0}^{1,\sigma}(t_0)$ in the spin propagation equation \eqref{eq:spin_propagation} to be in the spin state
\begin{equation}
\psi_i(\alpha) =
\begin{pmatrix}
 \cos \frac{\alpha}{2} \\
 \sin \frac{\alpha}{2}
\end{pmatrix}=
\begin{pmatrix}
 c_{0,0}^{1,0}(t_0) \\
 c_{0,0}^{1,1}(t_0)
\end{pmatrix}
\,.\label{eq:bloch_state}
\end{equation}
The form \eqref{eq:bloch_state} corresponds to a state on the Bloch sphere which points at some direction in the $x$-$z$ plane. On interaction of the laser with the electron, this quantum state gets diffracted by virtue of \eqref{eq:spin_propagation}, in our description. The resulting quantum state $c_{2,a''}^{1,\sigma''}(t)$ would be then of the form $(\eta \xi/2) \psi_i(\alpha + \phi)$, ie. rotated by the angle $\phi$ around the $y$-axis, with a reduced normalization, given by the factor $\eta \xi/2$. This rotation and change of normalization of the quantum space takes place for each index $a''$ with different values. The reader is reminded that the index $a''$ corresponds to the 5 different arrow directions of the diffracted wave packet, as illustrated in Fig. \ref{fig:Gaussian_beam}. Further details about spin rotations in the Kapitza-Dirac effect can be found in references \cite{ahrens_2012_phdthesis_KDE,ahrens_bauke_2013_relativistic_KDE} and generalizing concepts about possible other spin dynamics beyond a pure spin rotation are discussed in \cite{ahrens_2017_spin_filter,ahrens_2020_two_photon_bragg_scattering}.

While the rotation of the initial spin state $\psi_i$ by the spin propagation \eqref{eq:spin_propagation} with explicit form \eqref{eq:spin_propagation_matrix} is the description of the physical process, one may assign an even simpler picture to it, in the context of experimental detection. Assume, the electron was initially polarized along the $z$-axis
\begin{equation}
 c_{0,0}^{1,0}(t_0)=1\,,\quad c_{0,0}^{1,1}(t_0)=0\,.\label{eq:spin_up_initial_condition}
\end{equation}
This initial state corresponds to setting $\alpha=0$ in $\psi_i$ of Eq. \eqref{eq:bloch_state}. Then, with the spin propagation \eqref{eq:spin_propagation} and \eqref{eq:spin_propagation_matrix}, we can write the absolute square of the diffracted state as
\begin{subequations}%
\begin{align}%
|c_{2,a''}^{1,0}(t)|^2 &= \frac{\eta^2}{4} |\Braket{\mathds{1}|U}_{a''}|^2 = \frac{\eta^2 \xi^2}{4} \cos^2 \frac{\phi}{2}\label{eq:no_flip_probability}\\
|c_{2,a''}^{1,1}(t)|^2 &= \frac{\eta^2}{4} |\Braket{\mathds{\sigma}_y|U}_{a''}|^2 = \frac{\eta^2 \xi^2}{4} \sin^2 \frac{\phi}{2}\,.\label{eq:spin_flip_probability}
\end{align}%
\end{subequations}%
In essence, the coefficients $\Braket{\mathds{1}|U}_{a''}$ and $\Braket{\mathds{\sigma}_y|U}_{a''}$ can be associated with a diffraction probability $(\eta \xi/2)^2$ and a corresponding spin-flip probability \eqref{eq:spin_flip_probability} and no-flip probability \eqref{eq:no_flip_probability}. The probabilities are caused by a spin rotation around the $y$-axis during diffraction for each sub-diffraction order $a''$, ie. for each of the five arrows in Fig. \ref{fig:Gaussian_beam}.

\subsection{Numeric evaluation\label{sec:numeric_evaluation}}

Now we are ready for a numeric analysis of the quantum state propagation matrix $U$ in Eq. \eqref{eq:solution_perturbation_theory}, which we have cast into the form \eqref{eq:propagator_projections}. To give an orientation for the reader we first emphasize, how our work with an additional longitudinal polarization component from the Gaussian beam is extending previous calculations: The spin effect of the Kapitza-Dirac scattering in \cite{ahrens_bauke_2013_relativistic_KDE}, which is extended in this article, shows a flip of the electron spin. This spin flip is caused by a $\sigma_y$ expression of the electron spin propagation, as it is shown in \eqref{eq:spin_propagation_matrix} and corresponds to the $(z,z)$ index pair contribution in the perturbative propagation expression \eqref{eq:solution_perturbation_theory}. This $(z,z)$ contribution corresponds to a joint interaction of the transverse polarization component \eqref{eq:S_piture_interaction_transverse_vector_potential} for the potentials $V_{I}(t_{2})$ and, at the same time, $V_{I}(t_{1})$.

\begin{figure}%
    \includegraphics[width=0.5\textwidth]{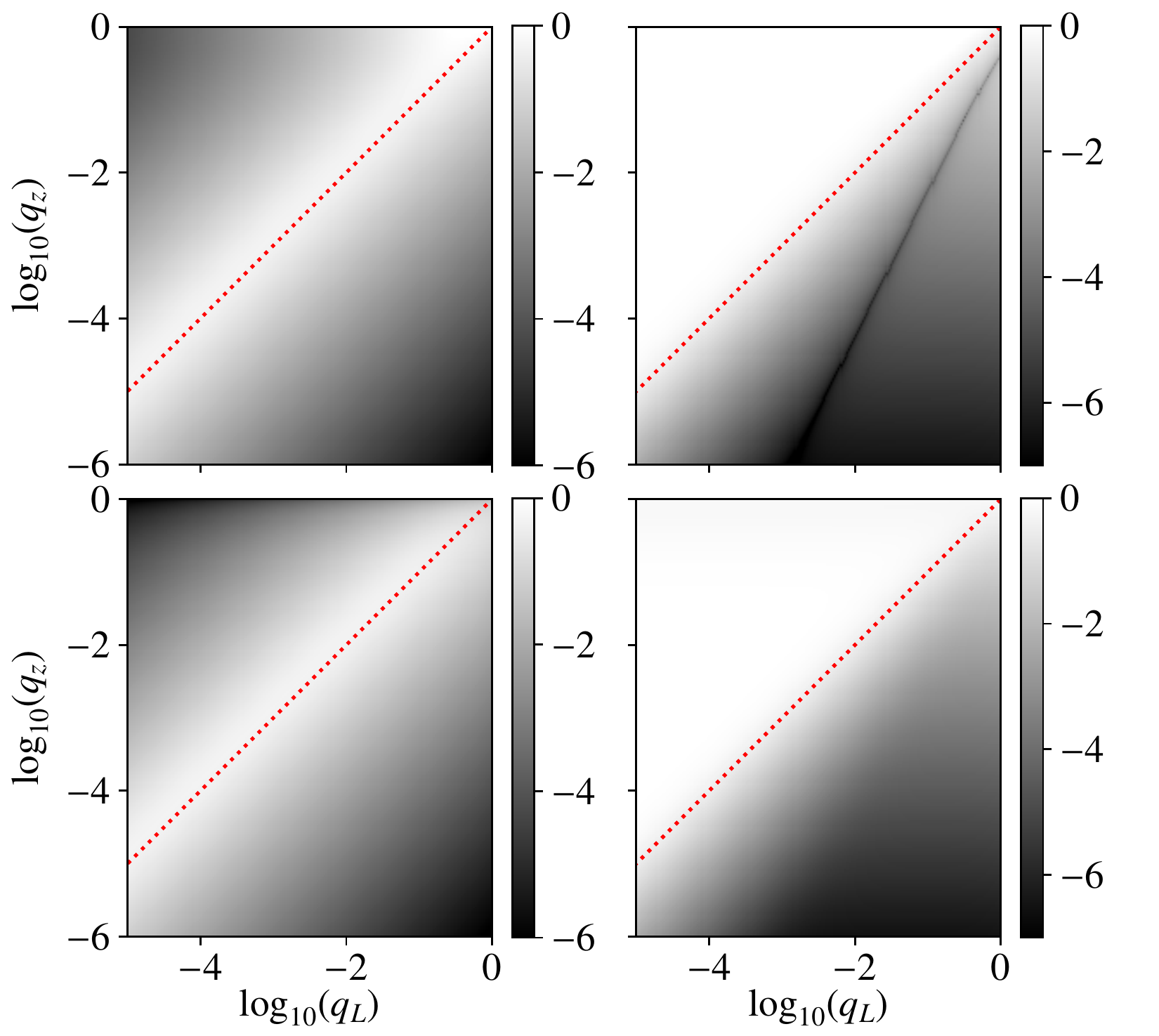}
    \caption{Amplitude of the functions $|\Braket{\mathds{1}|U}_{\pm 1}|$ and $|\Braket{\sigma_{y}|U}_{\pm 1}|$ as simultaneous functions of $q_L$ and $q_z$. The left (right) panels display the $|\Braket{\mathds{1}|U}|$ ($|\Braket{\sigma_{y}|U}|$) as implied by Eqs. \eqref{eq:propagator_projections}, respectively. The upper (lower) panels correspond to $a''=1$ ($a''=-1$), respectively. The red dotted line marks the location where $q_L = q_z$, at which $\epsilon=1/2$. Above $\epsilon=1/2$ (ie. in the upper left corner), the Gaussian beam approximation may be be considered as invalid. \label{Fig.1}}
\end{figure}

\begin{figure}%
    \includegraphics[width=0.5\textwidth]{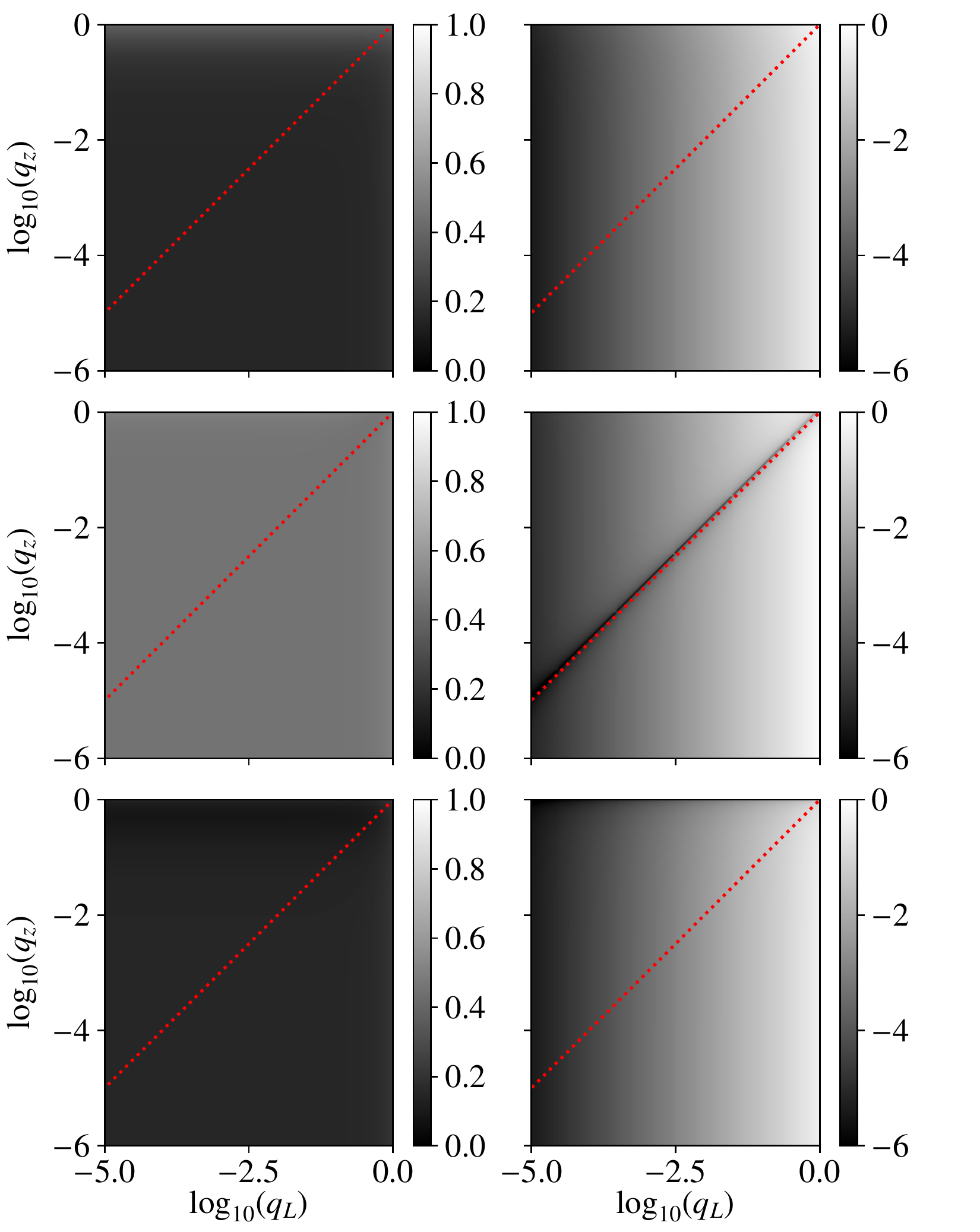}
    \caption{Amplitude of the functions $|\Braket{\mathds{1}|U}_{a''}|$ and $|\Braket{\sigma_{y}|U}_{a''}|$ for $a'' \in \{2, 0, -2\}$ as simultaneous functions of $q_L$ and $q_z$. This figure is similar to Fig. \ref{Fig.1}, except that the upper most panels correspond to $a''=2$, the middle panels correspond to $a''=0$ and the lower most panels correspond to $a''=-2$. The red dotted line marks the location, at which $\epsilon = 1/2$, see description in Fig. \ref{Fig.1} and in the main text.\label{Fig.2}}
\end{figure}

We are extending this term in our calculation by contributions which contain the action of the longitudinal polarization component \eqref{eq:S_piture_interaction_longitudinal_vector_potential} once (terms with index pairs $(x,z)$ or $(z,x)$) or even twice (terms with index pair $(x,x)$). For viewing modifications from the longitudinal beam component, we plot the amplitude of the spin preserving terms $\Braket{\mathds{1}|U}_{a''}$ and spin altering terms $\Braket{\sigma_{y}|U}_{a''}$ in Figs. \ref{Fig.1} and \ref{Fig.2} for different values of the final longitudinal electron momentum index $a''$, as a function of $k_L$ and $k_z$. Note, that according to our approach in \eqref{eq:momentum_vector} the $z$-component of the electron momentum of the final wave function is $a'' k_z + k_0$, where we set the $z$ momentum offset $k_0$ to the value $m$, consistent with previously considered scenarios in the references \cite{ahrens_2017_spin_non_conservation,ahrens_2020_two_photon_bragg_scattering}.

\begin{figure}%
    \includegraphics[width=0.5\textwidth]{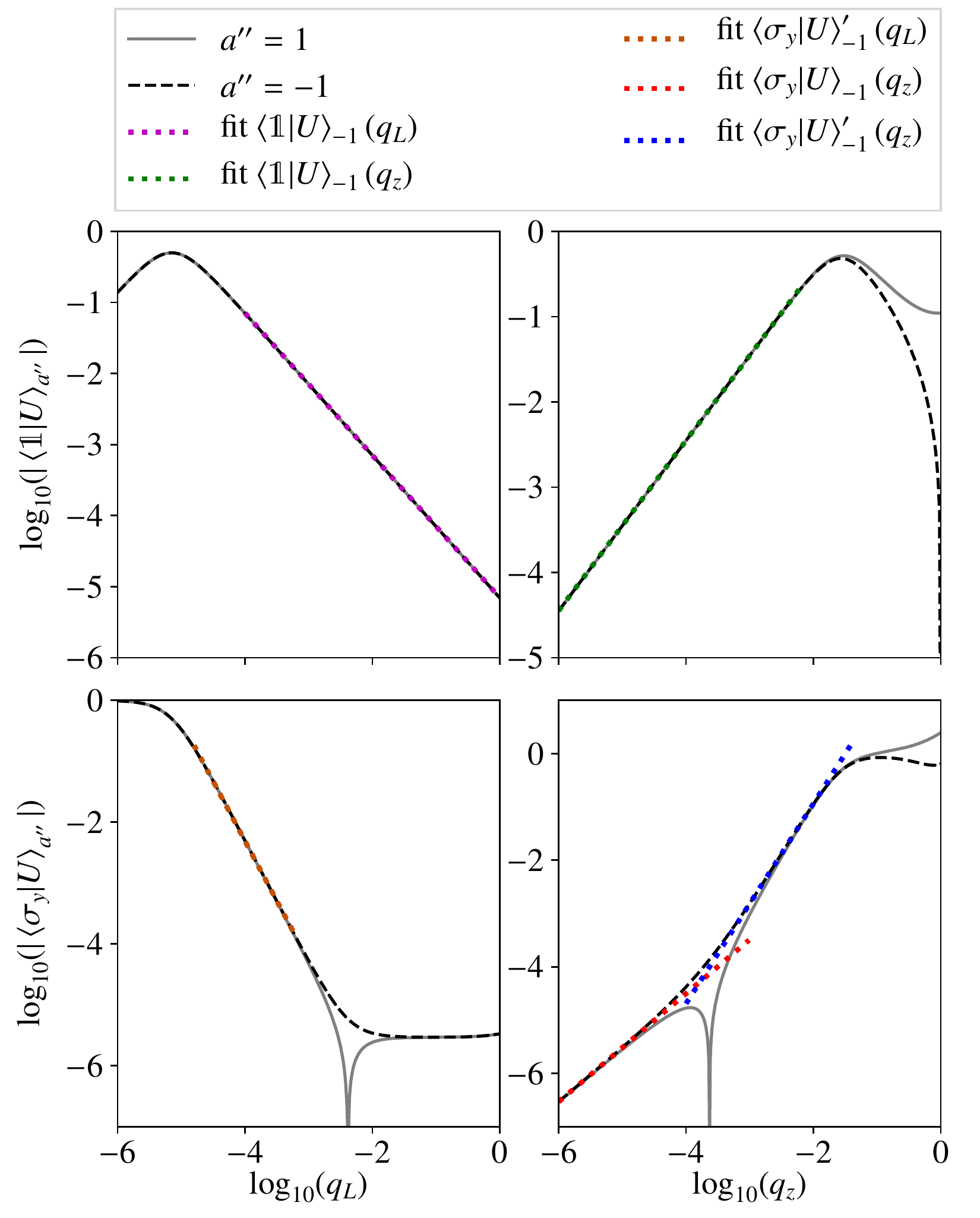}
    \caption{Amplitude of the functions $|\Braket{\mathds{1}|U}_{\pm 1}|$ and $|\Braket{\sigma_{y}|U}_{\pm 1}|$ as either a function of $q_L$ or $q_z$. The left panels show plots with varying $q_L$, where $q_z$ has the fixed value $10^{-5}$. Accordingly, the right panels show plots with varying $q_z$, where $q_L$ has the fixed value $2 \times 10^{-2}$. The colored, dashed lines are fits with the linear functions $h q_L + g$ (left panels) or $h q_z + g$ (right panels) to the linearly growing or dropping regions of the displayed functions, respectively. The slopes $h$ of the functions are listed in table \ref{tabel II}.\label{Fig.3}}
\end{figure}

\begin{figure}%
    \includegraphics[width=0.5\textwidth]{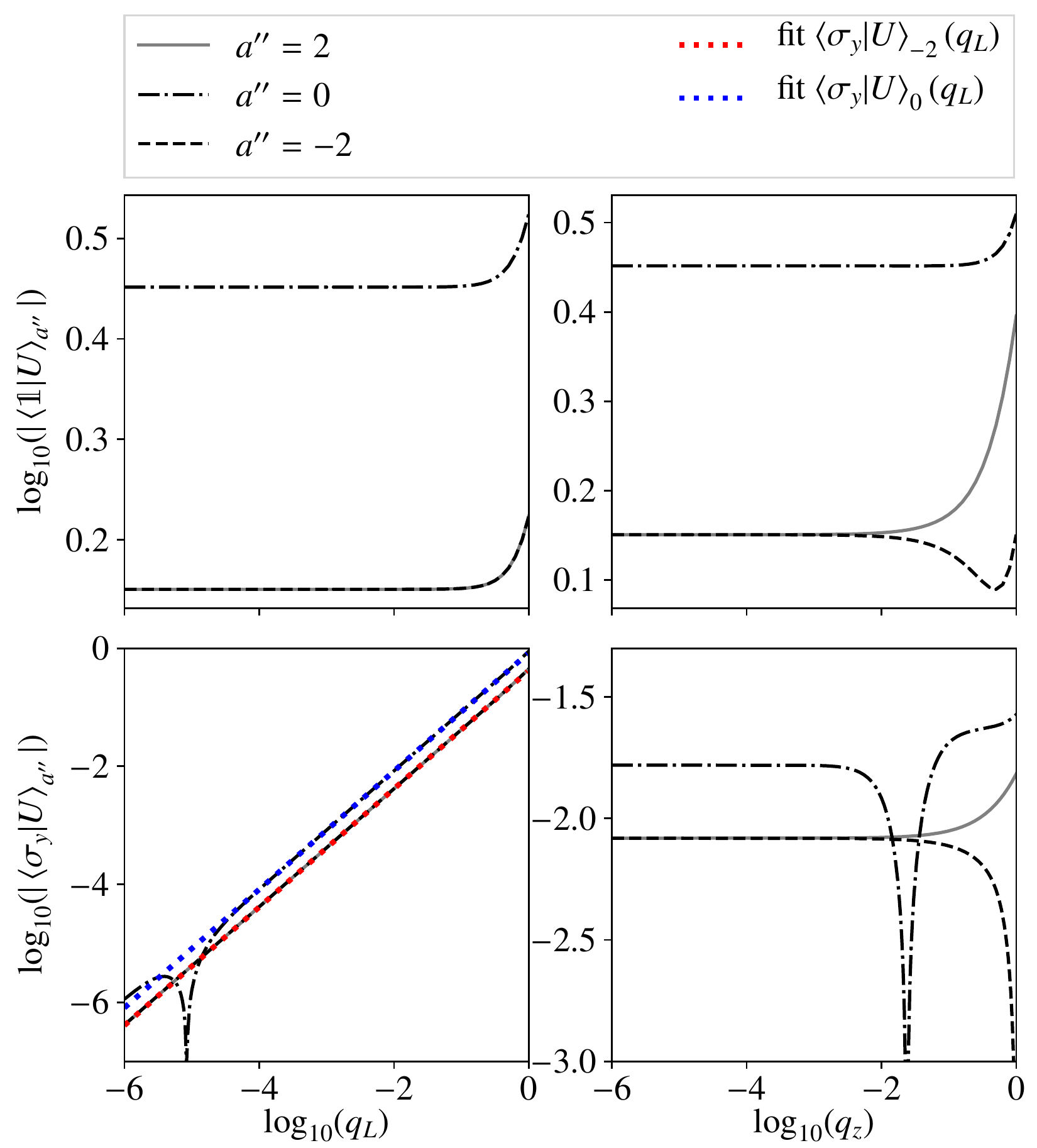}
    \caption{Amplitude of the functions $|\Braket{\mathds{1}|U}_{a''}|$ and $|\Braket{\sigma_{y}|U}_{a''}|$, with $a'' \in \{2, 0, -2\}$ as either a function of $q_L$ or $q_z$. Similarly to Fig. \ref{Fig.3} the left panels vary in $q_L$, with $q_z=10^{-5}$ and the right panel vary in $q_z$, with $q_L=2 \times 10^{-2}$. And correspondingly to Fig. \ref{Fig.3}, the colored, dashed lines in the lower left panel are linear fitting functions $h q_L + g$ of the linearly growing regions of the $\Braket{\sigma_{y}|U}$, with the slopes $h$ displayed in table \ref{tabel II}.\label{Fig.4}}
\end{figure}

It is more suitable to discuss the results in terms of the dimensionless variables
\begin{align}
 q_L = \frac{k_L}{m} \\
 q_z = \frac{k_z}{m}\,,
\end{align}
which are used in the following text and also in Figs. \ref{Fig.1} and \ref{Fig.2}. Note that in Fig. \ref{Fig.1} terms with \emph{one} longitudinal interaction are shown, which correspond to the index pairs $(x,z)$ or $(z,x)$, for which $a'' \in \{1,-1\}$, according to table \ref{tabel I}. On the contrary, in Fig. \ref{Fig.2} terms with \emph{two} longitudinal interactions are shown, corresponding to the index pair $(x,x)$, for which $a'' \in \{2,0,-2\}$. Note, that the solution for the Gaussian beam assumes, that the diffraction angle $\epsilon$ is small. Therefore, we have marked the location in Figs. \ref{Fig.1} and \ref{Fig.2} with a red dotted line, at which $\epsilon = 1/2$. Everything in the upper left corner, above this red dotted line corresponds to an unphysically large diffraction angle, at which the Gaussian beam approximation in powers of $\epsilon$ might be considered as invalid.
In order to investigate the functional behavior of $\Braket{\mathds{1}|U}$ and $\Braket{\sigma_{y}|U}$ more accurately, we plot them again in Figs. \ref{Fig.3} and \ref{Fig.4} as a function of either $q_L$ or $q_z$ in a line plot, instead the density plot of both variables in Figs. \ref{Fig.1} and \ref{Fig.2}. We set the fixed value of $q_z=10^{-5}$ in Fig. \ref{Fig.3} and $q_L=2 \times 10^{-2}$ Fig. \ref{Fig.4}. 

\subsection{Analysis of results\label{sec:analysis_of_results}}

We see in Figs. \ref{Fig.3} and \ref{Fig.4} a linear behavior of the functions $\Braket{\mathds{1}|U}$ and $\Braket{\sigma_{y}|U}$ over a vast area of the parameter range, which we approximate with linear fitting functions in the double-logarithmic plot. The linear function
\begin{equation}
 \log_{10}(|\braket{M|U}(\lambda)|) = h \lambda + g \label{eq:linear_fitting_function}
\end{equation}
with $\lambda = \log_{10}(q)$,  ($q$ is either $q_L$ or $q_z$ and $M$ is either $\mathds{1}$ or $\sigma_y$) can be written as power law
\begin{equation}
 |\braket{M|U}(q)| = q^h 10^g\,, \label{eq:power_law_function}
\end{equation}
where $q=10^\lambda$, such that the slope $h$ determines at which power $\Braket{M|U}$ is growing in $q$. We are listing the different slopes $h$ in table \ref{tabel II}.
\begin{table}
\caption{Slopes $h$ of the fitting functions in Figs. \ref{Fig.3} and \ref{Fig.4}. We show the parameter $h$ of the linear fitting functions $h \lambda + g$ in Eq. \eqref{eq:linear_fitting_function}, which are used in the double logarithmic plots of Fig. \ref{Fig.3} (above the horizontal line) and Fig. \ref{Fig.4} (below the horizontal line). The linear functions of the double logarithmic plots correspond to the power law $q^h 10^g$, corresponding to Eq. \eqref{eq:power_law_function}. For simplicity, we round the fitted slopes to integer numbers in the `approximation' column and use these simplified values for the exponents $\zeta$ and $\theta$ of the general functional form of $\braket{M|U}$ in Eq. \eqref{eq:propagator_scaling_law}. As a result, we obtain the specific functions \eqref{eq:scaling_longitudinal_1}, \eqref{eq:scaling_longitudinal_2}, \eqref{eq:rewritten_longitudinal_1} and \eqref{eq:rewritten_longitudinal_2}, according to the procedure as explained in the main text.\label{tabel II}}

\begin{tabular}{ l  l  l }			
  function$\qquad\qquad$ & slope $h$ & $\qquad\qquad$approximation\\
  \hline 
  $\Braket{\mathds{1}|U}_{-1}(q_L)\qquad$ & -1.0014 &$\hspace{7em}$ -1 \\
  $\Braket{\mathds{1}|U}_{-1}(q_z)\qquad$ & 0.998 & $\hspace{7.5em}$1\\
  $\Braket{\sigma_{y}|U}'_{-1}(q_L)\qquad$ & -1.96 &$\hspace{7em}$-2\\ 
  $\Braket{\sigma_{y}|U}_{-1}(q_z)\qquad$ & 1.015 & $\hspace{7.5em}$1\\
  $\Braket{\sigma_{y}|U}'_{-1}(q_z)\qquad$ & 1.88 & $\hspace{7em}$ 2 \\ \hline
  $\Braket{\sigma_{y}|U}_{-2}(q_L)\qquad$ & 1.0027 & $\hspace{7.5em}$1\\
  $\Braket{\sigma_{y}|U}_{0}(q_L)\qquad$ & 1.0031 & $\hspace{7.2em}$ 1
\end{tabular}
\end{table}

In order to obtain a simple scaling behavior of the functions $\Braket{M|U}$ on their linear range in the double-logarithmic plot, we denote them as
\begin{equation}
\Braket{M|U}=C q_L^{\zeta} q_z^{\theta}\,, \label{eq:propagator_scaling_law}
\end{equation}
where we take the approximated integer numbers in table \ref{tabel II} as the values for the corresponding powers $\zeta$ and $\theta$ of $q_L$ and $q_z$, respectively. The constant $C$ we obtain by probing the functions $\Braket{M|U}$ at specific value pairs $q_L$ and $q_z$ which we show in table \ref{tabel III} and solving Eq. \eqref{eq:propagator_scaling_law} for $C$. We obtain
\begin{subequations}%
\begin{alignat}{3}%
&\Braket{\mathds{1}|U}_{\pm 1}&&=&  &0.707 q_L^{-1} q_z\\
&\Braket{\sigma_{y}|U}_{1}    &&=& i&0.280 q_z \\
&\Braket{\sigma_{y}|U}_{-1}   &&=& i&0.305 q_z\\
&\Braket{\sigma_{y}|U}'_{1}   &&=&-i&0.460 q_L^{-2} q_z^{2}\\
&\Braket{\sigma_{y}|U}'_{-1}  &&=& i&0.497 q_L^{-2} q_z^{2}
\end{alignat}\label{eq:scaling_longitudinal_1}%
\end{subequations}%
for the expressions with a single longitudinal interaction, corresponding to the index pairs $(x,z)$ and $(z,x)$ and corresponding to Figs. \ref{Fig.1} and \ref{Fig.3}. Expressions with a double longitudinal interaction, corresponding to the index pair $(x,x)$ and Figs. \ref{Fig.2} and \ref{Fig.4} are resulting in
\begin{subequations}%
\begin{alignat}{3}%
&\Braket{\mathds{1}|U}_{\pm 2}&&=&  &1.414\\
&\Braket{\mathds{1}|U}_{0}    &&=& -&2.828\label{eq:longitudinal_2_id_a_0}\\
&\Braket{\sigma_{y}|U}_{\pm 2}&&=&-i&0.415 q_L\\
&\Braket{\sigma_{y}|U}_{0}    &&=& i&0.829 q_L\,.
\end{alignat}\label{eq:scaling_longitudinal_2}%
\end{subequations}%
We can further recast the expressions \eqref{eq:scaling_longitudinal_1} and \eqref{eq:scaling_longitudinal_2} by inserting relation \eqref{eq:k_z_k_L_relation},
written in the form
\begin{equation}
 q_z = 2 \epsilon q_L\,.
\end{equation}
and by multiplying with the $\Gamma$ factor from table \ref{tabel I}, resulting in
\begin{subequations}%
\begin{alignat}{3}%
&\Gamma_{\pm 1}\Braket{\mathds{1}|U}_{\pm 1}   &&=&\pm&0.606\epsilon_{}^{2}\\
&\Gamma_{1}\Braket{\sigma_{y}|U}_{1}           &&=&  i&0.240\epsilon_{}^{2} q_L \\
&\Gamma_{-1}\Braket{\sigma_{y}|U}_{-1}         &&=& -i&0.262\epsilon_{}^{2} q_L \\
&\Gamma_{1}\Braket{\sigma_{y}|U}'_{1}          &&=& -i&0.789\epsilon_{}^{3}\\
&\Gamma_{-1}\Braket{\sigma_{y}|U}'_{-1}        &&=& -i&0.853\epsilon_{}^{3}
\end{alignat}\label{eq:rewritten_longitudinal_1}%
\end{subequations}%
for the expressions which contain one longitudinal interaction. Here, we have substituted Eq. \eqref{eq:capital_epsilon} for the terms $\mathcal{E}$, which appear in the $\Gamma$ factor \eqref{prefactor} of table \ref{tabel I}. For expressions with two longitudinal interactions, in Eq. \eqref{eq:scaling_longitudinal_2} we obtain
\begin{subequations}%
\begin{alignat}{3}%
&\Gamma_{\pm 2}\Braket{\mathds{1}|U}_{\pm 2}  &&=&-i&0.260\epsilon_{}^{2}\\
&\Gamma_{0}\Braket{\mathds{1}|U}_{0}          &&=&-i&0.520\epsilon_{}^{2} \label{eq:rewritten_longitudinal_2_id_a_0}\\
&\Gamma_{\pm 2}\Braket{\sigma_{y}|U}_{\pm 2}  &&=&  &0.077\epsilon_{}^{2} q_L \\
&\Gamma_{0}\Braket{\sigma_{y}|U}_{0}          &&=&  &0.153\epsilon_{}^{2} q_L \,.
\end{alignat}\label{eq:rewritten_longitudinal_2}%
\end{subequations}%
Note, that we are using the final longitudinal diffraction order $a''$ as sub-index for the $\Gamma$ factors in Eqs. \eqref{eq:rewritten_longitudinal_1} and \eqref{eq:rewritten_longitudinal_2}.

\begin{table}
\caption{Specific function values of $\Braket{\mathds{1}|U}$ and $\Braket{\sigma_{y}|U}$ for the prefactor determination of the scaling approximation \eqref{eq:propagator_scaling_law}. The functions $\Braket{\mathds{1}|U}$ and $\Braket{\sigma_{y}|U}$ are evaluated at the parameter value pair $q_L$ and $q_z$, resulting in the column `value'. The values above the double lines correspond to Figs. \ref{Fig.1} and \ref{Fig.3} and below the double lines they correspond to Figs. \ref{Fig.2} and \ref{Fig.4}. Different types of functions are separated by single lines. The determined function values are used to solve the power law Eq. \eqref{eq:propagator_scaling_law} for the prefactor $C$, with corresponding values for $\zeta$ and $\theta$ from table \ref{tabel II}. The resulting functions are displayed in Eqs. \eqref{eq:scaling_longitudinal_1}, \eqref{eq:scaling_longitudinal_2}, \eqref{eq:rewritten_longitudinal_1} and \eqref{eq:rewritten_longitudinal_2}.\label{tabel III}}

\begin{tabular}{ c r r l }			
  function &  $\quad q_L\quad$ & $\quad q_z\quad$ & $\qquad\qquad$value\\
  \hline 
  $\Braket{\mathds{1}|U}_{1}$ & 2$\times 10_{}^{-2}$ & $\qquad 1 \times 10_{}^{-5}$ &$\qquad \phantom{-i}3.535\times 10_{}^{-4}$\\
  $\Braket{\mathds{1}|U}_{-1}$ & 2$\times 10_{}^{-2}$  & $\qquad 1 \times 10_{}^{-5}$ &$\qquad \phantom{-i}3.535\times 10_{}^{-4}$\\
  \hline 
  $\Braket{\sigma_{y}|U}_{1}$ & 2$\times 10_{}^{-2}$ & $\qquad1 \times 10_{}^{-5}$ &$\qquad\phantom{-} i2.804\times 10_{}^{-6}$\\ 
  $\Braket{\sigma_{y}|U}_{-1}$ & 2$\times 10_{}^{-2}$ & $\qquad1 \times 10_{}^{-5}$ &$\qquad \phantom{-} i3.054\times 10_{}^{-6}$\\
  \hline
  $\Braket{\sigma_{y}|U}'_{1}$ &2$\times 10_{}^{-2}$ & $\qquad 6 \times 10_{}^{-3}$&$\qquad -i4.140\times 10_{}^{-2}$ \\
  $\Braket{\sigma_{y}|U}'_{-1}$ & 2$\times 10_{}^{-2}$ & $\qquad6 \times 10_{}^{-3}$&$\qquad \phantom{-} i4.474\times 10_{}^{-2}$\\
  \hline  \hline
  $\Braket{\mathds{1}|U}_{2}$ & 1$\times 10_{}^{-3}$  & $\qquad 1 \times 10_{}^{-5}$&$\qquad \phantom{-i} 1.414\times 10_{}^{0}$\\
  $\Braket{\mathds{1}|U}_{-2}$ & 1$\times 10_{}^{-3}$  & $\qquad 1 \times 10_{}^{-5}$&$\qquad \phantom{-i} 1.414\times 10_{}^{0}$\\
  $\Braket{\mathds{1}|U}_{0}$ & 1$\times 10_{}^{-3}$  & $\qquad 1 \times 10_{}^{-5}$&$\qquad  - \phantom{i}2.828\times 10_{}^{0}$\\
  \hline
  $\Braket{\sigma_{y}|U}_{2}$ & 2$\times 10_{}^{-2}$ & $\qquad1 \times 10_{}^{-5}$ &$\qquad -i8.285\times 10_{}^{-3}$\\
  $\Braket{\sigma_{y}|U}_{-2}$ & 2$\times 10_{}^{-2}$ & $\qquad1 \times 10_{}^{-5}$ &$\qquad -i8.285\times 10_{}^{-3}$\\
  $\Braket{\sigma_{y}|U}_{0}$ & 2$\times 10_{}^{-2}$ & $\qquad1 \times 10_{}^{-5}$ &$\qquad \phantom{-} i1.657\times 10_{}^{-2}$
\end{tabular}
\end{table}

\section{Discussion and Conclusion\label{sec:discussion_and_conclusion}}

The resulting values in tables \ref{tabel II} and \ref{tabel III} and expressions in Eqs. \eqref{eq:scaling_longitudinal_1} till \eqref{eq:rewritten_longitudinal_2} are correction terms for the interaction of the longitudinal laser polarization component with the electron. In the introduction, the question was posed how a longitudinal polarization component from beam focusing is influencing the spin-dynamics of the Kapitza-Dirac effect. In order to answer this question, the expressions \eqref{eq:scaling_longitudinal_1} till \eqref{eq:rewritten_longitudinal_2} need to be compared with the purely transverse polarization component interaction, which corresponds to the $(z,z)$ index pair. For this index pair, we find a linear scaling of $\Braket{\sigma_{y}|U}$ of $q_L$ with power $\zeta=1$ and the value $\Braket{\sigma_{y}|U}$ at $q_L=2.0\times10^{-2}$ has the value $i 2\times10^{-2}$, in consistency with our previous expressions in reference \cite{ahrens_2020_two_photon_bragg_scattering} \footnote{We point out that different normalizations in the bi-spinor definitions have been used in this work, as compared to reference \cite{ahrens_2020_two_photon_bragg_scattering}}. The laser-electron interaction can be constructed such that $\Braket{\mathds{1}|U}$ can vanish completely for the $(z,z)$ index pair, see footnote [89] in reference \cite{ahrens_2017_spin_non_conservation} and the statement around Eq. (15) in reference \cite{ahrens_2020_two_photon_bragg_scattering}. Since the $(z,z)$ contribution does not contain any beam waist dependent longitudinal interaction components, $\Braket{\mathds{1}|U}$ and $\Braket{\sigma_{y}|U}$ are independent of $q_z$. Therefore, for the form in Eq. \eqref{eq:propagator_scaling_law} we have $\zeta=1$ and $\theta=0$ for $\Braket{\sigma_{y}|U}$, and in analogy to Eqs. \eqref{eq:scaling_longitudinal_1} and \eqref{eq:scaling_longitudinal_2} obtain
\begin{subequations}%
\begin{alignat}{3}%
&\Braket{\mathds{1}|U}_{0}&&=& &0\label{eq:zero_order_spin_not_changing_without_gamma}\\
&\Braket{\sigma_{y}|U}_{0}&&=&i&q_L\label{eq:zero_order_spin_changing_without_gamma}
\end{alignat}\label{eq:zero_order_without_gamma}%
\end{subequations}%
and correspondingly with $\Gamma_{z,z} = -1$
\begin{subequations}%
\begin{alignat}{3}%
&\Gamma_{0}\Braket{\mathds{1}|U}_{0}&&=&  &0\label{eq:zero_order_spin_not_changing}\\
&\Gamma_{0}\Braket{\sigma_{y}|U}_{0}&&=&-i&q_L\,,\label{eq:zero_order_spin_changing}
\end{alignat}\label{eq:zero_order}%
\end{subequations}%
in analogy to Eqs. \eqref{eq:rewritten_longitudinal_1} and \eqref{eq:rewritten_longitudinal_2}, where we again substituted $a''=0$ for the index pair $(z,z)$ in the sub-index of $\Gamma$.

The spin changing electron-laser interaction without longitudinal contribution in Eq. \eqref{eq:zero_order_spin_changing} corresponds to the situation, in which beam focusing is completely neglected. It therefore is independent of the diffraction angle $\epsilon$. In contrast to that, interaction contributions with a longitudinal component in Eqs. \eqref{eq:rewritten_longitudinal_1} and \eqref{eq:rewritten_longitudinal_2} all scale at least with power 2 in $\epsilon$. In other words, the influence of longitudinal field components from laser beam focusing on the investigated, spin-dependent effect in Kapitza-Dirac scattering gets arbitrary low for arbitrary low beam foci, within the approximations which have been made in this article.

Besides these general considerations, it is also interesting to give an estimate at what values of the interaction parameters $q_L$ and $\epsilon$ the influence from the longitudinal polarization component begins to matter for the spin-dynamics. Since there are multiple sub-diffraction orders $a''$ with spin-preserving and spin-flipping contributions (9 different terms in Eqs. \eqref{eq:rewritten_longitudinal_1} and \eqref{eq:rewritten_longitudinal_2}), which have a partially different scaling behavior, it is reasonable to concentrate on contributions which scale with the smallest power in the small quantities $q_L$ and $\epsilon$, for an estimation. Most notably might be the contribution in Eq. \eqref{eq:rewritten_longitudinal_2_id_a_0}, which is proportional to the spin-preserving $2\times2$ identity and has the final longitudinal diffraction order $a''=0$. With $a''=0$, this contribution is located in the same point in momentum space as the interaction term \eqref{eq:zero_order} of an interaction without longitudinal contribution. Therefore both terms are physically indistinguishable after the interaction, by any means. Furthermore, \eqref{eq:rewritten_longitudinal_2_id_a_0} only scales with $\epsilon^2$ and is therefore one of the largest contributions from the longitudinal interactions. Eq. \eqref{eq:rewritten_longitudinal_2_id_a_0} is therefore our candidate for estimating the longitudinal influence on the spin dynamics in the following calculation. More specifically, we use the equivalent form  \eqref{eq:longitudinal_2_id_a_0} of Eq. \eqref{eq:rewritten_longitudinal_2_id_a_0} for clarity of the calculation, in the following. Similarly to the considerations in section \ref{sec:spin_propagation}, one obtains the approximate probability for observing a diffraction without spin-flip
\begin{equation}
 |c_{2,a''}^{1,0}(t)|^2 = \left[ \frac{\epsilon}{2}\frac{2.828}{e}\frac{q^2 A_0^2}{4}  (t-t_0) \right]^2 \approx \left[ \frac{\epsilon}{2}\frac{q^2 A_0^2}{4}  (t-t_0) \right]^2
\label{eq:leading_longitudinal}
\end{equation}
from Eq. \eqref{eq:longitudinal_2_id_a_0}, when inserted into \eqref{eq:spin_propagation_matrix}, with initial condition \eqref{eq:spin_up_initial_condition}. Eq. \eqref{eq:leading_longitudinal} stems from an interaction of the electron with a longitudinal polarization component and needs to be compared with the terms \eqref{eq:zero_order_without_gamma}, which do not involve interactions with the longitudinal component. The spin-preserving term \eqref{eq:zero_order_spin_not_changing_without_gamma} is zero and is therefore not of relevance for an estimation for the leading interaction contribution. The remaining spin-flipping term \eqref{eq:zero_order_spin_changing_without_gamma} inserted in Eq. \eqref{eq:spin_propagation_matrix} with initial condition \eqref{eq:spin_up_initial_condition} gives the spin-flip probability
\begin{equation}
 |c_{2,a''}^{1,1}(t)|^2 = \left[ q_L \frac{q^2 A_0^2}{4}  (t-t_0) \right]^2
 \,.\label{eq:leading_pure_transverse}
\end{equation}
According to our explanations from above, spin-dynamics in the Kapitza-Dirac effect are getting influenced by a longitudinal polarization component, when the diffraction probability \eqref{eq:leading_longitudinal} is getting on the order of magnitude of the probability \eqref{eq:leading_pure_transverse}. Thus, setting both probabilities equal results in the scaling law
\begin{equation}
 q_L = \frac{\hbar k_L}{m c} = \frac{\lambda_C}{\lambda} = \frac{\epsilon^2}{2}\,,\label{eq:beam_focus_relevance}
\end{equation}
which tells at which wavelengths $\lambda$ and diffraction angles $\epsilon$ the contributions from longitudinal interaction components are turning into non-negligible amplitudes. The reduced Planck constant $\hbar$, the vacuum speed of light $c$, the Compton wavelength $\lambda_C$ and the wavelength of the laser light $\lambda = 2 \pi/k_L$ are written out explicitly in Eq. \eqref{eq:beam_focus_relevance} for clearness.

There are two interesting light frequencies (photon energies) for possible applications. One photon energy is the hard X-ray regime where 10\,keV roughly correspond to $q_L=2\times10^{-2}$, which is the value which is mainly under study in this paper. For this photon energy we obtain $\epsilon=0.2$, which implies beam foci on the order of
\begin{equation}
 w_0 = \frac{\lambda}{2 \pi \epsilon} = 96\,\textrm{pm}\,.\label{eq:beam_focus}
\end{equation}
The other interesting photon energy is 2 \,eV of red light with 620\,nm, corresponding approximately to $q_L=4\times 10^{-6}$. For this parameter we have $\epsilon = 2.8\times 10^{-3}$ and Eq. \eqref{eq:beam_focus} yields $35\,\mu\textrm{m}$ for the corresponding laser beam focus. We therefore conclude, that longitudinal fields from beam focusing are not expected to have significant influence on the spin dynamics of the investigated scenario of a spin altering Kapitza-Dirac effect with a hard X-ray standing light wave. In the optical regime however, the influence of longitudinal fields might be of relevance for the electron spin dynamics.

\section{Outlook\label{sec:outlook}}

The main motivation for our study was to answer, whether a longitudinal polarization component from beam focusing has an influence on spin dynamics in the Kapitza-Dirac effect. In this context, we have only accounted for the  transverse spacial dependence of the longitudinal component, but did not account for the transverse spacial dependence of the transverse polarization component, which however, one would expect to scale with at least $\epsilon^2$. In other words, this first investigation could still be improved into a study which is consistent up to order $\epsilon^2$, within our plane wave approximation of the potentials. Along this line, one might raise the question, whether the rough approximation of the potentials in Eqs. \eqref{eq:Gaussain_beam_transverse_vector_potential} and \eqref{eq:Gaussain_beam_longitudinal_vector_potential} are sufficiently accurate for solid statements at all. It is possible, to solve the relativistic quantum dynamics of the Dirac equation by exact numeric solutions, for example with the Fourier-transform split-operator method \cite{Grobe_1999_FFT_split_operator_method,bauke_2011_GPU_acceleration_FFT_split_operator}. With this type of more exact solution approach, a systematic parameter study as it is done in this article would be more difficult, but there would be less doubts about a possible oversimplification of the problem.

One interesting detail, which is not accounted for in the spin-\emph{changing} electron dynamics of this work is the investigation of spin-\emph{dependent} diffraction \cite{McGregor_Batelaan_2015_two_color_spin,dellweg_mueller_2016_interferometric_spin-polarizer,dellweg_mueller_extended_KDE_calculations,ahrens_2017_spin_filter,ebadati_2018_four_photon_KDE,ebadati_2019_n_photon_KDE,ahrens_2020_two_photon_bragg_scattering}. While the diffraction pattern in spin-dependent scattering would depend on the initial electron spin state, the diffraction pattern would be completely independent of the initial spin for the case of spin-changing dynamics. In other words, spin-dependent dynamics allows for the implementation of a Stern-Gerlach type of experiment, while spin-changing dynamics does not have this capability. However, till now, there is no purely linear field configuration known for implementing spin-dependent diffraction for the interaction with a single laser pulse and with only a two-photon-interaction in the Bragg regime. This implies complications for theoretical investigations, because the transverse spacial dependence of the longitudinal laser polarization component turns two-dimensional for any type of elliptical polarization, resulting in the necessity of a quantum simulation in three dimensions. Three dimensional solutions of the Dirac equation, in turn, are challenging, though not impossible \cite{Fu_2019_3D_dirac_spin_solution}, because one needs to numerically resolve the fast oscillations of the electron wave function in the complex plane which are implied by the mass term of the Dirac equation. One could of course think of solving the Schr\"odinger equation plus spin coupling terms for this problem, but beside the necessity of numerical time propagation techniques of operators which are neither diagonal in position space nor diagonal in momentum space, one would also encounter the question about which relativistic corrections from the Foldy-Wouthuysen transformations of the Dirac equation are of relevance for the electron spin dynamics. Configurations are known for which the plain Pauli equation is not enough for describing the system correctly \cite{bauke_ahrens_2014_spin_precession_1,bauke_ahrens_2014_spin_precession_2}.

\begin{acknowledgments}
The work was supported by the National Natural Science Foundation of China (Grants No. 11975155 and No. 11935008) and the Ministry of Science and Technology of the People's Republic of China (Grants No. QN20200213003, No. 2018YFA0404803 and No. 2016YFA0401102).
\end{acknowledgments}
\vspace{0.5 cm}

\appendix

\section{Field of Gaussian beam from reference \cite{Quesnel_1998_gaussian_beam_coulomb_gauge}\label{sec:gaussian_vector_potential}}

Reference \cite{Quesnel_1998_gaussian_beam_coulomb_gauge} presents the Gaussian beam in form of the electric fields
\begin{subequations}%
\begin{align}%
E_{x}=&E_{0}\frac{w_{0}}{w}\exp\left(-\frac{r^2}{w^2}\right)\sin\left(\phi_{G}\right)
\label{x_direction_electric_field}\\
E_{z}=&2E_{0}\epsilon\frac{xw_{0}}{w^2}\exp\left(-\frac{r^2}{w^2}\right)\cos\left(\phi_{G}^{(1)}\right)
\label{z_direction_electric_field}
\end{align}\label{eq:Gaussian_beam_electric_field}%
\end{subequations}%
with phases
\begin{subequations}%
\begin{align}%
\phi_{G}=&\omega_{0} t-k_0 z+\tan^{-1}\left(\frac{z}{z_{R}}\right)-\frac{zr^2}{z_{R}w^2}-\phi_{0}\label{eq:phase_phi_G_electric}\\
\phi_{G}^{(1)}=&\phi_{G}+\tan^{-1}\left(\frac{z}{z_{R}}\right)
\end{align}\label{eq:phases_electric_field}%
\end{subequations}%
and beam waist along the $z$-direction
\begin{equation}
w(z)=w_{0}\sqrt{1+\frac{z^2}{z_{R}^2}}\,.
\label{eq:beam_waist}
\end{equation}
Though the Gaussian beam is given as electric field, we follow the argument in reference \cite{Quesnel_1998_gaussian_beam_coulomb_gauge}, where the electric field components are recast to the vector field components. This procedure is justified, by pointing out that \eqref{eq:Gaussian_beam_electric_field} is obtained from the Maxwell-Poisson equation $\vec \nabla \cdot \vec E = 0$ and the scalar wave equation in vacuum
\begin{equation}
 \left( \Delta - \frac{1}{c^2} \frac{\partial^2}{\partial t^2} \right) \psi =0\,. \label{eq:scalar_wave_equation}
\end{equation}
One can see a formal equivalence to the Coulomb gauge condition $\vec \nabla \cdot \vec A = 0$ in combination with \eqref{eq:scalar_wave_equation} and therefore, in accordance with reference \cite{Quesnel_1998_gaussian_beam_coulomb_gauge}, we apply the substitution $E_i \rightarrow A_i$, $i \in \{x,y,z\}$ for the field components and $E_0 \rightarrow A_0$ for the field amplitude in the Gaussian beam \eqref{eq:Gaussian_beam_electric_field}. Furthermore, we already use $k_0$ for the electron momentum but $k_L$ for the laser's wave number and also a plain omega $\omega$ for the laser frequency in this article. Correspondingly, we also substitute $k_0 \rightarrow k_L$ and $\omega_0 \rightarrow \omega$ in Eq. \eqref{eq:phase_phi_G_electric}. Additionally, since we desire a beam propagating along the $x$-axis, we perform a space rotation on the potential \eqref{eq:Gaussian_beam_electric_field} and phase \eqref{eq:phases_electric_field} by applying the coordinate substitution $z\rightarrow x, x\rightarrow -z, y\rightarrow y$ for the vectorial quantities. Also, since the Kapitza-Dirac effect is based on the interaction with two beams, which are propagating in opposite directions, we denote a second, counterpropagating beam configuration from a subsequent rotation by 180$\degree$, by imposing the coordinate substitution $x\rightarrow -x, y\rightarrow -y, z\rightarrow z$. Applying the mentioned substitutions to \eqref{eq:Gaussian_beam_electric_field} and \eqref{eq:phases_electric_field} results in Eqs. \eqref{eq:vector_field} and \eqref{eq:Gaussian_beam_phase} in section \ref{sec:gaussian_beam_introduction}.

\bibliography{bibliography}

\phantom{\cite{Fradkin_Gitman_Shvartsman_1991_Quantum_Electrodynamics_with_Unstable_Vacuum,woellert_2015_pair_creation_tunneling,woellert_2016_multi_pair_states,lv_bauke_2017_multi_pair_creation}}

\end{document}